\documentclass[12pt]{article}
\pdfoutput=1
\usepackage{draft}
\usepackage{hyperref}
\usepackage{graphicx,color,subfig}
\usepackage{cite}
\usepackage{mciteplus}
\usepackage{skak}
\usepackage{empheq}
\usepackage{tikz}
\usepackage{booktabs}
\usepackage{siunitx}
\usepackage{bbm}
\usepackage{makecell}
\usepackage{float}
\usepackage{dsfont}

\usepackage{bm}
\usepackage{braket}

\newcommand{\ba}{\begin{aligned}}
\newcommand{\ea}{\end{aligned}}

\def\be{\begin{equation}}
\def\ee{\end{equation}}

\begin{document}

\unitlength = 1.8mm

\begin{titlepage}

\begin{center}

\hfill \\
\hfill \\
\vskip 1cm

\title{Entanglement Entropy and Modular Hamiltonian of free fermion with deformations on a torus }

\author{\large{Song He$^{1,2,}$, Zhang-Cheng Liu$^{1,}$, Yuan Sun$^{1}$}}
{\normalsize\it $^1$Center for Theoretical Physics and College of Physics, Jilin University,\\ Changchun 130012, People's Republic of China
\\$^{2}$Max Planck Institute for Gravitational Physics (Albert Einstein Institute),\\ Am M\"uhlenberg 1, 14476 Golm, Germany}

\email{hesong@jlu.edu.cn, zhangchenliuxt@gmail.com, sunyuan@jlu.edu.cn}
\end{center}

\abstract{In this work, we perturbatively calculate the modular Hamiltonian to obtain the entanglement entropy in a free fermion theory on a torus with three typical deformations, e.g., $T\bar{T}$ deformation, local bilinear operator deformation, and mass deformation. For $T\bar{T}$ deformation, we find that the leading order correction of entanglement entropy is proportional to the expectation value of the undeformed modular Hamiltonian. As a check, in the high/low-temperature limit, the entanglement entropy coincides with that obtained by the replica trick in the literature. Following the same perturbative strategy, we obtain the entanglement entropy of the free fermion vacuum state up to second-order by inserting a local bilinear operator deformation in a moving mirror setting. In the uniformly accelerated mirror, the first-order and second-order correction of entanglement entropy vanishes in the late time limit. For mass deformation, we derive the entanglement entropy up to first-order deformation and comment on the second-order correction.}
\vfill

\end{titlepage}

\tableofcontents

\section{Introduction}
Entanglement is crucial to exploring the information encoded in the quantum theory. One of the important tools to demonstrate the entanglement between complementary spatial regions is entanglement entropy. The system is divided into the subsystem $A$ and the complement subsystem $B$. The entanglement entropy between $A$ and $B$ is defined as the von Neumann entropy as
\begin{equation}\label{e1l}
S_A=-\text{Tr}_A[\rho_A\log\rho_A ],
\end{equation}
where $\rho_A$ is the reduced density matrix of the subsystem $A$.
Usually, the direct analysis of the entanglement entropy in the logarithmic form as (\ref{e1l}) is complex. In practice, one may use the replica trick to compute the entanglement entropy \cite{Calabrese:2004eu}. In this approach, one firstly calculates the  R\'{e}nyi entropy on an n-fold cover of a manifold. Then the entanglement entropy could be obtained by taking the $n\to 1$ limit of the R\'{e}nyi entropy. In the viewpoint of holography, the entanglement entropy corresponds to the area of extremal surface in the bulk spacetime which is known as the Ryu-Takayanagi(RT) formula \cite{Ryu:2006bv, Hubeny:2007xt}.

The entanglement entropy can also be evaluated from the expectation value of modular Hamiltonian  $\mathcal{K}_A$ of the subsystem $A$ which by definition is related to reduced density matrix as
\begin{equation}\label{modular}
\rho_A=\frac{e^{-2\pi\mathcal{K}_A}}{\text{Tr}_Ae^{-2\pi\mathcal{K}_A}}.
\end{equation}
Modular Hamiltonian can be used to calculate the relative entropy of excited state \cite{Lashkari:2015dia,Sarosi:2017rsq}. It plays a crucial role in proving of the first law of entanglement \cite{Alcaraz:2011tn,Bhattacharya:2012mi,Nozaki:2013vta,Guo:2013aca,Allahbakhshi:2013rda,Blanco:2013joa,Faulkner:2013ica} and the averaged
null energy condition \cite{Faulkner:2016mzt}. The gravity duality of modular Hamiltonian becomes a effective tool to reconstruct the bulk operator in \cite{Jafferis:2014lza,Jafferis:2015del,Faulkner:2018faa}. In terms of the modular Hamiltonian, the entanglement entropy can be written as
\begin{equation}\label{es}
S_A=2\pi\langle\mathcal{K}_A\rangle+\log(\text{Tr}_Ae^{-2\pi\mathcal{K}_A}).
\end{equation}

In general, the modular Hamiltonian is non-local, and the explicit expression is difficult to write down. Nevertheless, there are a few cases where the modular Hamiltonian is known analytically in field theory. From the result of Bisognano and Wichmann \cite{Bisognano:1975ih,Bisognano:1976za}, the modular Hamiltonian is the boost generator for the half-plane $x>0$ of the flat Minkowski space. For a conformal field theory (CFT), the authors of \cite{Casini:2011kv, Wong:2013gua} obtained the modular Hamiltonian of a spherical region by the conformal mapping from the thermal state. When the subsystem is conformally equivalent to the annulus, the modular Hamiltonian is constructed in \cite{Cardy:2016fqc}. In the case of chiral fermion in $1+1$ dimension, the modular Hamiltonian is provided for the gaussian state by using the resolvent \cite{Cardy:2016fqc, Casini:2009vk,Mintchev:2020uom, Fries:2019ozf}.

Furthermore, it is nontrivial to explore the entanglement entropy and modular Hamiltonian for more generic entangling surfaces and states, even employing the perturbative approach \cite{Rosenhaus:2014woa}. When the entangling surface between the subregions is disturbed, it was suggested in \cite{Balakrishnan:2017bjg} that the modular Hamiltonian's shape dependence is needed for proving the quantum null energy condition. One can refer to \cite{Faulkner:2015csl, Allais:2014ata, Rosenhaus:2014zza} for studying the shape dependence of the entanglement entropy of vacuum state in both dual CFTs and the gravity side. The entanglement entropy and modular Hamiltonian of excited state for a ball shape region are investigated in \cite{Sarosi:2017rsq} in which the authors give a correspondence from both sides of the holographic duality. The evaluation of modular Hamiltonian of excited state can be found in \cite{Balakrishnan:2020lbp, Arias:2020qpg, Lashkari:2018oke}.

It is a nontrivial temptation to understand the generic structure of modular Hamiltonian in generic quantum field theory, even in a perturbative sense. In recent years, much attention has been paid to the $T\bar{T}$ deformed quantum field theory which keeps the integrability properties of the undeformed theory \cite{Cavaglia:2016oda, Smirnov:2016lqw}. From holographic viewpoint,  the $T\bar{T}$ deformed CFTs dual to the cutoff $AdS_3$ spacetime\cite{McGough:2016lol}, which is refined by \cite{Kraus:2022mnu} recently. Under such duality, the holographic entanglement entropy can be computed by RT surface in finite-cutoff geometry \cite{McGough:2016lol,Donnelly:2018bef,Chen:2018eqk, Ota:2019yfe,Donnelly:2019pie,Allameh:2021moy,Setare:2022qls,Kraus:2022mnu}.\footnote{ There is an alternative holographic dictionary which stated that the $T\bar{T}$ deformed CFT is dual to the $AdS_3$ with mixing boundary condition ~\cite{Guica:2020uhm,Kraus:2021cwf,He:2021bhj}.} In field theory side, the entanglement entropy of vacuum state and local excited states in deformed theories has been investigated in \cite{Jeong:2019ylz}\cite{He:2019vzf}\cite{He:2020qcs}\cite{Asrat:2020uib}\cite{Cardona:2022cmh}. Besides the $T\bar{T}$ deformation, studying the free field theory with mass deformation is also interesting. In \cite{Rosenhaus:2014zza}ï¼Œ authors showed the entanglement entropy of free fermions and scalars \cite{Ageev:2022kpm} with mass deformation matches the results from the holographic calculation,  which suggests that the free field theory is useful for understanding the holography. Further, one can obtain the modular Hamiltonian and its flow by turning on an local operator in the path integral formalism in \cite{Arias:2020qpg, Rangamani:2015agy}, as well as the time evolution of entanglement entropy for free fermion under the local bilinear operator deformation \cite{Nozaki:2015mca}.

In this work, we focus on the modular Hamiltonian in free fermion field theory on a torus with three kinds of deformation mentioned above, i.e., $T\bar{T}$ deformation, mass term deformation, and local bilinear operator deformation. In the case of  $T\bar{T}$ deformation, we study the chiral fermion on a torus  \cite{Fries:2019acy} whose entanglement entropy can be calculated from the modular Hamiltonian and the technical details of developed in \cite{Rosenhaus:2014ula,moosa}. Further, following similar construction in \cite{Rangamani:2015agy,Nozaki:2015mca}, we turn on a local bilinear operator deformation in moving mirror setup \cite{Akal:2021foz} to mimic hawking radiation \cite{WanMokhtar:2018lwi} under the disturbance of the locally external source. Finally, we investigate the modular Hamiltonian from its local term in the free fermion on a torus with a mass term deformation \cite{Herzog:2013py}.

The remainder of this manuscript is organized as follows. In Sec. \ref{a}, the modular Hamiltonian on a torus and the conventions are introduced. Next, the modular Hamiltonian of a single interval on a cylinder and an interval at the end of a semi-infinite line is computed. To study the entanglement entropy of $T\bar{T}$ deformed fermions in Sec. \ref{b}, the entanglement entropy of free fermion with a $T\bar{T}$ deformation for a single interval on a torus is firstly analyzed from the modular Hamiltonian in Sec. \ref{eottt}. In this case, the modular Hamiltonian's local and non-local parts are evaluated. Next, the entanglement entropy correction of free fermion with a $T\bar{T}$ deformation for a half-line is evaluated in Sec. \ref{e}.  Then, we obtain the entanglement entropy on a torus of a local bilinear operator deformed fermions in Sec. \ref{localdf}. In Sec. \ref{f}, the entanglement entropy for moving mirror of chiral fermion with a local bilinear operator is studied. Following a similar method, we derive entanglement entropy on a torus of mass deformed fermions in Sec. \ref{mee}. Finally, the conclusions is presented in Sec. \ref{g}.

\section{Entanglement entropy and modular Hamiltonian}\label{a}
In this section, we establish the notations and conventions for the modular Hamiltonian of free chiral fermion on a torus, i.e., finite size and finite temperature system. At the same time, we review the derivation of modular Hamiltonians on a cylinder and a half-line that we will need for our analysis.

It was found in \cite{Casini:2009vk} that the modular Hamiltonian of the chiral fermion field consists of two parts: the local modular Hamiltonian and the bi-local modular Hamiltonian. The latter appears for multi-interval cases on a plane. The situation is more complicated on a torus, where the modular Hamiltonian would develop both local and bi-local parts even for a single interval  \cite{Fries:2019ozf}.

To demonstrate the properties of the Gaussian state,\footnote{Gaussian state is a state of a Hamiltonian which is quadratic in the creation or annihilation operator.} the author of \cite{Inch} gave an example in the fermionic hopping model. The Hamiltonian of the system satisfies $\hat{H}=-\sum_{n,m}t_{n,m}c^\dagger_nc_m$ where $t_{n,m}$ is the hopping amplitude between nearest sites and $c_{i,j}$ is the fermion field of the model.

Study a subregion $\mathcal{M}$ with the notation $i,j$ label the sites. For the state which is Slater determinant, the high correlation function $C_{i,j}$ can be factorized through the one point function as $C_{i,j}=tr(\rho c^\dagger_ic_j)$. According to the Wick's theorem, when the reduced matrix $\rho$ is exponential of the free-fermion operator, one has
 $\rho=\mathcal{K}e^{-\mathcal{H}}$, where $\mathcal{K}$ is the normalization constant and $\mathcal{H}=\sum_{i,j}H_{i,j}c^\dagger_ic_j$.
 of the chiral fermion field $\psi$, the reduced density matrix is \cite{Blanco:2019cet,Inch} also Gaussian. For a subystem restricted to a single interval $A=(a,b)$ on a torus, the modular Hamiltonian takes the form
\begin{equation}\label{ker}
\mathcal{K}_A=\int_Adxdy\psi^\dagger(x)k_A(x,y)\psi(y), \quad x,y\in A
\end{equation}
where $k_A(x,y)$ is the kernel which is related with the two-point correlation function $G_A(x,y)$ through $k_A=-\log(G_A^{-1}-1)$.\footnote{Here $G_A(x,y)\equiv \langle\psi(x)\psi^\dagger(y)\rangle$  is the two-point function of the chiral fermions with $x,y\in A$.}
It was shown in \cite{Fries:2019ozf} that the modular Hamiltonian of chiral fermion on a torus  can be separated into a local term  and a bi-local term as
\begin{equation}\label{hmer}
\mathcal{K}_A=\mathcal{K}_A^{local}+\mathcal{K}_A^{bi-local}.
\end{equation}
As our study involves the system in Minkowski spacetime and Euclidean spacetime, we will introduce the corresponding modular Hamiltonian and stress tensor separately.

In the Minkowski spacetime with the coordinate $(x,t)$, the local modular Hamiltonian for chiral fermion is constructed by an integral as
\begin{equation}\label{localh2}
\tilde{\mathcal{K}}_A^{local}=\int_A\beta(x)T_{tt}(x,t)dx,
\end{equation}
where $\beta(x)=\frac{2\pi\beta}{2\pi+\beta\partial_x\log\Omega_A( x|\tau)}$ is the entanglement temperature with $\Omega_A(x|\tau)=-\frac{\vartheta_1( x-a|\tau)}{\vartheta_1( x-b|\tau)}$.\footnote{The periods of a torus is taken to be $1,\tau=i\beta$. $\vartheta_1$ is the Jacobi-$\vartheta$ function shown in the App. \ref{9.4}. And define $\gamma_t=\begin{pmatrix}0&1\\1&0\end{pmatrix},\gamma_x=\begin{pmatrix}0&-1\\1&0\end{pmatrix}$.} The $tt$-component of stress tensor is $T_{tt}(x,t)=\frac{i}{2}[\psi^\dagger(x,t)\gamma_t\gamma_x\partial_x\psi(x,t)-\partial_x\psi^\dagger(x,t)\gamma_t\gamma_x\psi(x,t)]
$ with 2-component spinor $\psi(x,t)=(\psi_1(x+t),\psi_2(x-t))$, or
\begin{equation}\label{St12}
T_{tt}(x,t)=\frac{i}{2}\Big(\psi_1^\star(x+t)\partial_x\psi_1(x+t)-\partial_x\psi_1^\star(x+t)\psi_1(x+t)\Big)-\frac{i}{2}\Big(\psi_2^\star(t-x)\partial_x\psi_2(t-x)-\partial_x\psi_2^\star(t-x)\psi_2(t-x)\Big).
\end{equation}
 For chiral fermion on a torus corresponding to spin sectors $\nu=2$ (real periodic, complex antiperiodic) and $\nu=3$ (doubly antiperiodic)\cite{Fries:2019ozf},  the bi-local modular Hamiltonian is
\footnote{ For the cases $\nu=1,4$, the related Green's function is unbounded. There is no good expression for the entanglement entropy in the resolvent approach pointed out in\cite{Fries:2019acy}}
\begin{equation}\label{localbi2}
\tilde{\mathcal{K}}^{bi-local}_{\pm}=\int_A\sum_{k\in\mathbb{Z} \backslash \{0\}}(\pm 1)^k\tilde{\beta}(x,x_k(x))\psi^\dagger(x,t)\psi(x_k(x),t)\delta\Big(x-x_k+\beta \frac{1}{2\pi(b-a)}\log\frac{\Omega_A(x|\tau)}{\Omega_A(x_k|\tau)}-k\Big)dx,
\end{equation}
where the symbol $+,-$ correspond to the spin sector $\nu=2, 3$. And the points $x_k$ satisfies the  equation below
\begin{equation}\label{kl}
x-x_k+\beta \frac{1}{2\pi(b-a)}\log\frac{\Omega_A(x|\tau)}{\Omega_A(x_k|\tau)}-k=0.
\end{equation}
Here the bi-local entanglement temperature defines as
\begin{equation}
\tilde{\beta}(x,y)=\frac{i\pi}{(b-a)\sinh\frac{1}{2(b-a)}\log\frac{\Omega_A(x|\tau)}{\Omega_A(y|\tau)} }.
\end{equation}
In Euclidean signature, the authors of \cite{Blanco:2019cet} studied the entanglement entropy and modular Hamiltonian for chiral fermions on a torus from the images method. In the limit $\beta\to\infty$, the modular Hamiltonian in \cite{Blanco:2019cet} is consistent with   (\ref{localh2}) and (\ref{localbi2})  after the analytic continuation of imaginary to real-time.

In the present work, the modular Hamiltonian in Euclidean signature is obtained by the analytic continuation  $\tilde{\tau}\to it$ of (\ref{localh2}), (\ref{localbi2}) and (\ref{St12}). Then the stress tensor reads
\begin{equation}\label{St1}
T_{zz}=\frac{i}{2}\big(\psi^*(z)\partial\psi(z)-\partial\psi^*(z)\psi(z)\big)
\end{equation}
with the complex coordinate  $z=x+i\tilde\tau,\bar{z}=x-i\tilde\tau$.\footnote{Following \cite{DiFrancesco:1997nk}, we introduce
$
T=-2\pi T_{zz}, \bar{T}=-2\pi T_{\bar{z}\bar{z}}, \Theta=-2\pi T_{z\bar{z}}$
.}The local and bi-local modular Hamiltonian in Euclidean spacetime are respectively
\begin{equation}\label{localh}
\mathcal{K}_A^{local}=-\int_A \beta(x)T_{\tilde{\tau}\tilde{\tau}}(x,\tilde{\tau})dx,
\end{equation}
\begin{equation}\label{localbi}
\mathcal{K}^{bi-local}_{\pm}=-\int_A\sum_{k\in\mathbb{Z} \backslash \{0\}}(\pm 1)^k\tilde{\beta}(x,x_k(x))\psi^\dagger(x,\tilde{\tau})\psi(x_k(x),\tilde{\tau})\delta\Big(x-x_k+\beta \frac{1}{2\pi(b-a)}\log\frac{\Omega_A(x|\tau)}{\Omega_A(x_k|\tau)}-k\Big)dx.
\end{equation}

\subsection{Single interval on a cylinder}\label{rain}
    Before studying the modular Hamiltonian on a torus, two specific examples are given to show the construction of the modular Hamiltonians by conformal mapping \cite{{Cardy:2016fqc}}, which will be used in Sec. \ref{albb}. In the first case, we study a cylinder that is equivalent to a torus  at the zero temperature limit $\beta\to\infty$. The modular Hamiltonian for a single interval $V=\{(-R,R), \ell=2R\}$ takes the following form \cite{{Cardy:2016fqc}}
\begin{equation}\label{vmodular}
\mathcal{K}_V=\int_C\frac{(T_{zz}+\frac{c}{24\pi}\{f(z);z\})}{f'(z)}dz+\int_{\bar{C}}\frac{(\bar{T}_{\bar{z}\bar{z}}+\frac{c}{24\pi}\{f(\bar{z});{\bar{z}}\})}{f'(\bar{z})}d\bar{z}.
\end{equation}
This expression corresponds to the local modular Hamiltonian~(\ref{localh}) discussed in previous section. $C$ is the intersection of the spatial region $V$ with a constant time slice and $\{f(z);z\}$ is the Schwarzian derivative with $f(z)=\log \left(\frac{e^{2\pi i z}-e^{-2\pi i R}}{e^{2\pi i R}-e^{2\pi i z}} \right )$. $c$ represents the central charge. \footnote{Here we have $-T_{\tilde{\tau}\tilde{\tau}}(\tilde\tau,x)=T_{zz}+\bar{T}_{\bar{z}\bar{z}}$.}
For the static case evaluated hereafter, we will take a Euclidean time slice at $\tilde{\tau}=0$. Then $C=\bar{C}$, and the modular Hamiltonian (\ref{vmodular}) is simplified to \footnote{The extra minus sign in front of the integral comes from the fact that the Hamiltonian density in Minkowski signature is different from the euclidean one by a minus sign.}
 \begin{equation}\label{1mh}
\mathcal{K}_V=-\int_C\frac{T_{\tilde{\tau}\tilde{\tau}}(x)}{f'(x)}dx+\frac{c}{12\pi}\int_C\frac{\{f(x);x\}}{f'(x)}dx.
\end{equation}
For a cylinder of circumference $L$, the one-point function for stress tensor is \cite{DiFrancesco:1997nk}
\begin{equation}\label{stress}
\langle T_{\tilde{\tau}\tilde{\tau}}\rangle=\frac{\pi c}{6 L^2}.
\end{equation}
Plugging (\ref{stress}) and the function $f(z)$ into (\ref{1mh}), the modular Hamiltonian becomes \footnote{Details of the calculation is demonstrated in the App. \ref{9.1}.}
\begin{equation}
\langle\mathcal{K}_V\rangle=\frac{c}{12\pi}\log\Big(\frac{L}{\pi\epsilon}\sin{\frac{\pi l}{L}}\Big)+O(1).
\end{equation}

\subsection{Single interval at the end of a semi-infinite line}\label{storm}
As for the other example, investigate a spacial subregion $A=(-R,0)$ at the end of semi-infinite line $B=(-\infty,0)$. Following the similar procedure in the above subsection, one can construct the modular Hamiltonian of the interval $A$ through the conformal transformation $f(z)=\log\frac{R+z}{R-z}$. Plugging $f(z)$ into~(\ref{1mh}), then it is shown that
\begin{equation}
\langle\mathcal{K}_A\rangle=\frac{c}{12\pi}\int_A \frac{R}{R^2-x^2}dx,
\end{equation}
 where we have used the one-point function of stress tensor on the half-plane $\langle T(z)\rangle=0$ \cite{McAvity:1995zd}. Then the modular Hamiltonian is
\begin{equation}
\langle\mathcal{K}_A\rangle=\frac{c}{12\pi}\log{\frac{2R}{\epsilon}}.
\end{equation}

\section{Entanglement entropy of $T\bar{T}$ deformed fermions}\label{b}
In this section, we would like to investigate the entanglement entropy for free chiral fermion with $T\bar{T}$ deformation perturbatively. Both finite (Sec. \ref{eottt}) and infinite (Sec. \ref{e}) system are taken into consideration.

Let us begin with  deriving the entanglement entropy of a system under general deformation denoted as $\mathcal{T}$  by employing modular Hamiltonian. The perturbed action upto the first order in coupling constant $\lambda$ reads \footnote{Here the coordinate transformation is chosen to be $d^2x=\frac{i}{2}d^2z$.}
\begin{equation}\label{actiond}
\mathcal{I}_\lambda=\mathcal{I}_{CFT}-\lambda\mathcal{T}.
\end{equation}
For the $T\bar{T}$ deformation, we have $\mathcal{T}\equiv\mathcal{T}_T=\frac{i}{2}\int_{\mathcal{M}} d^2zT(z)\bar{T}(\bar{z})$. In the same setting,  the $T\bar{T}$ deformed correlators on a torus were evaluated by using the perturbative approach in \cite{He:2020udl}.
The entanglement entropy of the subsystem  $A$ can be expanded as\footnote{The superscript $(i)$ represents the $i$-th order correction of expectation value. For simplicity, we drop the expansion parameter and one can read the expansion order from the power of coupling constant $\lambda$.}
\begin{equation}\label{ET}
S_A(\lambda)=S_A^{(0)}+\lambda\frac{dS_A(\lambda)}{d\lambda}\Bigg|_{\lambda=0}+\frac{\lambda^2}{2}\frac{d^2S_A(\lambda)}{d^2\lambda}\Bigg|_{\lambda=0}+....,
\end{equation}
From eq.~(\ref{modular}), the entanglement entropy can be written in path integral formalism as \cite{moosa}
\begin{equation}\label{el}
S_A(\lambda)=\frac{1}{Z_\lambda}\int D\psi e^{-\mathcal{I}_\lambda}\Big(2\pi\mathcal{K}_A(\lambda)+\log(tr_Ae^{-2\pi\mathcal{K}_A(\lambda)})\Big),
\end{equation}
where $Z_\lambda$ is the partition function of the system.
It is derived in \cite{Rosenhaus:2014ula,moosa} that the first-order derivative of the entanglement entropy is \footnote{Please refer to App. \ref{9.b} for a detailed derivation.}
\begin{equation}\label{see1}
\frac{dS_A(\lambda)}{d\lambda}=\langle2\pi\mathcal{T}\mathcal{K}_A(\lambda)\rangle_\lambda-\langle\mathcal{T}\rangle_\lambda\langle2\pi\mathcal{K}_A(\lambda)\rangle_\lambda.
\end{equation}
Plugging (\ref{see1}) into (\ref{ET}), the entanglement entropy is
\begin{align}\label{deed}
S_A(\lambda)&=S_A^{(0)}+2\pi\lambda\Big(\langle\mathcal{T}\mathcal{K}_A(\lambda)\rangle_\lambda-\langle\mathcal{T}\rangle_\lambda\langle\mathcal{K}_A(\lambda)\rangle_\lambda\Big)\Big|_{\lambda=0}+O(\lambda^2)&\nonumber
\\&=S_A^{(0)}+\Delta S_A(\lambda),
\end{align}
where $\Delta S_A(\lambda)$ is the correction of the entanglement entropy of deformed theory.
Then the leading order correction of the entanglement entropy $S_A^{(1)}$ gives the first law of entanglement entropy \cite{Bhattacharya:2012mi,Blanco:2013joa,Rosenhaus:2014ula}ï¼Œ
\begin{equation}\label{fmh}
S_A^{(1)}=2\pi\lambda\Big(\langle\mathcal{T}\mathcal{K}_A(\lambda)\rangle_\lambda-\langle\mathcal{T}\rangle_\lambda\langle\mathcal{K}_A(\lambda)\rangle_\lambda\Big)\Big|_{\lambda=0}.
\end{equation}

\subsection{Single interval on a torus}\label{eottt}
\subsubsection{Local  modular Hamiltonian  }\label{albb}
As mentioned before the modular Hamiltonian of chiral fermions for a single interval $A=(a,b)$ on a torus contains local and bi-local terms. We first investigate the contribution from the local modular Hamiltonian to the entanglement entropy of $T\bar{T}$ deformed chiral fermions.
From the (\ref{fmh}), we have
\begin{equation}\label{dh}
{S_A^{local}}^{(1)}=2\pi\lambda\Big(\langle\mathcal{T}_T\mathcal{K}_A^{local}(\lambda)\rangle_\lambda-\langle\mathcal{T}_T\rangle_\lambda\langle\mathcal{K}_A^{local}(\lambda)\rangle_\lambda\Big)\Big|_{\lambda=0}.
\end{equation}
To calculate (\ref{dh}), we first introduce the partition function and correlation function for chiral fermions on a torus which will be needed for analysis.
The partition function on a torus for general CFT is \footnote{$L_0$ is the zero mode operator of the energy density.}
\begin{equation}
Z=Tr(q^{L_0-c/24}{\bar{q}}^{(\bar{L}_0-c/24)}),\quad q=e^{2\pi i \tau}.
\end{equation}
For the free fermion with spin structure $\nu$, the partition function on a torus is factorized into holomorphic and antiholomorphic parts as \cite{DiFrancesco:1997nk}
\begin{equation}
Z_\nu=Z'_\nu\bar{Z}'_\nu,\quad Z'_\nu(\tau)=\Big(\frac{\vartheta_\nu(\tau)}{\eta(\tau)}\Big)^{1/2}.
\end{equation}
where $\eta(\tau)$ is the Dedekind function, $\vartheta_\nu$ is Jacobi-$\nu$ function. The two point correlation function of chiral fermion with spin secotr $\nu$ satisfies
\begin{align}
\langle\psi^\star(z)\psi(w)\rangle_\nu&=\frac{1}{2\pi i}P_{\nu}(z-w),\nonumber\\
\langle\bar{\psi}^\star(\bar{z})\bar{\psi}(\bar{w})\rangle_\nu&=\frac{1}{2\pi i}\bar{P}_{\nu}(\bar{z}-\bar{w}),\quad \nu=2,3,4
\end{align}
where \begin{equation}\label{wf}
P_\nu(z)=\sqrt{P(z)-e_{\nu-1}}=\frac{\vartheta_\nu(z|\tau)\partial_z\vartheta_1(0|\tau)}{\vartheta_1(z|\tau)\vartheta_\nu(0|\tau)}.
\end{equation}
Here $P(z)$ is the Weierstrass $P$ function defined in the App. \ref{9.4}. One can apply point-splitting regularization  \cite{He:2020cxp} to compute the following one-point functions
\begin{align}
\langle\partial\psi^\star(z)\psi(z)\rangle_\nu=\lim_{w\to z}(\langle\partial\psi^\star(z)\psi(w)\rangle_\nu+(z-w)^{-2})=\frac{1}{2\pi i}\partial P_\nu(0)=-\frac{1}{2\pi i}e_{\nu-1}.
\end{align}
{To fix the second term of (\ref{dh}), one can apply the  $\langle T(z)\bar{T}(\bar{z})\rangle_0$\footnote{ $\langle ...\rangle_0$ represents the expectation value in the undeformed theory. } given by \cite{He:2020udl}.
\begin{equation}
\langle T(z)\bar{T}(\bar{z})\rangle_0=-(2\pi i)^2\frac{1}{Z}\partial_\tau\partial_{\bar{\tau}}Z,
\end{equation}
Then one can obtain}
\begin{equation}\label{ettbar}
\langle\mathcal{T}_T\rangle_0=\frac{i}{2}\int_{\mathcal{M}}\langle T(z)\bar{T}(\bar{z})\rangle_0d^2z=-i(2\pi i)^2\tau\frac{1}{Z}\partial_\tau\partial_{\bar{\tau}}Z.
\end{equation}
As indicated in (\ref{localh}), the local modular Hamiltonian is expressed as the integral of the energy density which can be separated into holomorphic and antiholomorphic parts, i.e. $T(w)+\bar{T}(\bar{w})$. Accordingly, we separated the local modular Hamiltonian into two parts
\begin{equation}
\langle\mathcal{K}_A^{local}\rangle_0=\langle\mathcal{K}_A^{local}\rangle^h_0+\langle\mathcal{K}_A^{local}\rangle^{\bar{h}}_0
\end{equation}
with
\begin{equation}\label{hl}
\langle\mathcal{K}_A^{local}\rangle^h_0= i\partial_{\tau}\ln Z \int_A\beta(x)dx,\quad
\langle\mathcal{K}_A^{local}\rangle^{\bar{h}}_0= -i\partial_{\bar{\tau}}\ln\bar{Z}\int_A\beta(x) dx.
\end{equation}
Here we used $\langle T(w)\rangle=-2\pi i\partial_{\tau}\ln Z$,  $\langle\bar{T}(\bar{w})\rangle=2\pi i\partial_{\bar{\tau}}\ln\bar{ Z}$ \cite{He:2020udl} and (\ref{localh}).

Plugging the modular Hamiltonian defined in~(\ref{localh}) into the first term of ~(\ref{dh}), we obtain
\begin{equation}\label{ttbarh}
\langle\mathcal{T}_T\mathcal{K}_A^{local}(\lambda)\rangle_0^h=\int_{\mathcal{M}}\int_A\beta(w)\langle T_{ww}(w)T(z)\bar{T}(\bar{z})\rangle_0d^2zdw,
\end{equation}
where we have $\langle\mathcal{T}_T\mathcal{K}_A^{local}\rangle_0^h=\langle\mathcal{T}_T\mathcal{K}_A^{local}\rangle_0^{\bar{h}}$. The correlator in the integral eq. (\ref{ttbarh}) equals  \cite{He:2020udl}
 \begin{equation}\label{3T}
\langle T_{ww}(w)T(z)\bar{T}(\bar{z})\rangle_0=\frac{4\pi^2i\partial^2_{\tau}\partial_{\bar{\tau}} Z}{Z}-\frac{ic}{12}\partial^2P(w-z)\partial_{\bar{\tau}}\ln Z+(P(w-z)+2\eta_1)
\frac{4\pi\partial_\tau\partial_{\bar{\tau}} Z}{ Z},
\end{equation}
where $\eta_1=\zeta(1/2)$, and $\zeta(z)$ is the zeta function. And we set $Z=Z_\nu$ for convenience hereafter. Thus we obtain
\begin{equation}\label{ttbarh2}
\langle\mathcal{T}_T\mathcal{K}_A^{local}(\lambda)\rangle_0
=\int_A\beta(x) \Big(\frac{8\pi^2\tau \partial^2_{\tau}\partial_{\bar{\tau}} Z}{Z}+\frac{8\pi^2\partial_\tau\partial_{\bar{\tau}}Z}{Z} \Big) dx,
\end{equation}
where we have used the integrals in eq. (\ref{c8}) and eq. (\ref{c9}).
Plugging (\ref{ttbarh2}), (\ref{ettbar}) and~(\ref{hl}) into~(\ref{dh}), the leading order correction of entanglement entropy from the local modular Hamiltonian is
\begin{align}\label{Dlocal}
\frac{{S_A^{local}}^{(1)}}{2\pi}&=\lambda\int_A\beta(x)\Big(\frac{8\pi^2\tau \partial^2_{\tau}\partial_{\bar{\tau}} Z}{Z}+\frac{8\pi^2\partial_\tau\partial_{\bar{\tau}} Z}{Z}-4\pi^2\tau\frac{1}{Z}\partial_\tau\partial_{\bar{\tau}}Z(\partial_{\tau}\ln Z-\partial_{\bar{\tau}}\ln\bar{Z}\Big)dx.&
\end{align}
Note that the terms in the bracket on the RHS are coordinate-independent. Comparing the (\ref{Dlocal}) with (\ref{hl}), the correction of the entanglement entropy is proportional to the expectation value of local modular Hamiltonian for the undeformed fermions that
\begin{align}\label{loee}
{S_A^{local}}^{(1)}&=2\pi\lambda\langle\mathcal{K}_A^{local}(\lambda)\rangle_0\times\frac{1}{( i)\partial_{\bar\tau}\ln Z -( i)\partial_{\tau}\ln Z }\Big(\frac{8\pi^2\tau \partial^2_{\tau}\partial_{\bar{\tau}} Z}{Z}+\frac{8\pi^2\partial_\tau\partial_{\bar{\tau}}Z}{Z}\nonumber\\
&-4\pi^2\tau\frac{1}{Z}\partial_\tau\partial_{\bar{\tau}}Z(\partial_{\tau}\ln Z-\partial_{\bar{\tau}}\ln\bar{Z})\Big),
\end{align}
where the undeformed modular Hamiltonian is $\langle{\mathcal{K}_A}^{local}\rangle_0= i(-\partial\tau\ln Z+\partial_{\bar{\tau}}\ln\bar{Z})\int_A\beta(x)dx$. This local modular Hamiltonian can be calculated numerically. For the spin sector $\nu=3$ in the low-temperature limit $\beta\to\infty$, the local modular Hamiltonian is simplified to $2\pi\langle{\mathcal{K}_A}^{local}\rangle_0= {S_A^{(local)}}^{(0)}-\log(Z)$ \footnote{ By using the resolvent in \cite{Fries:2019acy}, the spin-independent entanglement entropy for a single interval $(a,b)$ in CFT on a torus is
$
{S_A^{local}}^{(0)}=\frac{1}{6}\Big(\log|\vartheta_1(b-a|\tau)|-\log|\vartheta_1(\epsilon|\tau)|\Big)
$} according to~(\ref{es}), where the bi-local part of modular Hamiltonian is zero \cite{Fries:2019ozf}.

By taking the large size limit in the space direction of the torus, the leading order correction of entanglement entropy in (\ref{loee}) can be reduced to the results on a cylinder. To show this, one can make use of the modular transformation $\tau\to -\frac{1}{\tau}$ and take the ``high temperature'' limit $\beta\to 0$.
 In the situation $\tau\to 0$, we obtain\footnote{Full demonstration can be found in App. \ref{9.2} and \ref{9.21}}
\begin{equation}\label{slocal}
{S_A^{local}}^{(1)}=\frac{32\lambda\pi^3\partial_\tau\partial_{\bar{\tau}} Z}{Z}\int_A\beta(x)  dx,
\end{equation}
According to~(\ref{ettbar}),  the stress tensor of a cylinder with Ramond boundary condition in the time direction is
{\begin{equation}\label{cst}
\langle T(z)\bar{T}(\bar{z})\rangle_0=\Big(\frac{\pi^2 c}{6\beta^2}\Big)^2 .
\end{equation}}
Plugging~(\ref{temp}),~(\ref{cst}),~(\ref{ettbar}) and~(\ref{slocal}) into~(\ref{loee}), the leading order correction to the entanglement entropy is
\begin{equation}\label{bqe}
\lim_{\beta\to 0} S_A^{local(1)}(\lambda)=\frac{\lambda l \pi^4 c^2}{9\beta^3}\coth\frac{\pi l}{\beta}.
\end{equation}
It matches the entanglement entropy in $T\bar{T}$ deformed CFT on a cylinder (2.22) in \cite{Chen:2018eqk}.

To compare the entanglement entropy on a torus with the result on a cylinder at zero temperature, we evaluate the torus corresponding to the low-temperature limits $\beta\to\infty$.
Using the Wick's theorem,\footnote{Please refer to \cite{He:2020cxp} for related analysis.} the entanglement entropy from ~(\ref{Dlocal}) becomes
\begin{align}
\lim_{\beta\to\infty}{S_A^{local}}^{(1)}&=-i\pi\lambda\lim_{\beta\to\infty}\int_A\beta(x)\int_{\mathcal{M}}\Big(\frac{\bar{\partial}\bar{P}_\nu(0)(\partial {P_\nu(z-x))}^2}{\pi}-\frac{\bar\partial \bar{P}_\nu(0)\partial^2 P_\nu(z-x)P_\nu(z-x)}{\pi}\Big)dxd^2z&\nonumber\\
&=2\lambda\lim_{\beta\to\infty}\int_A\beta(x) \bar{\partial}\bar{P}_\nu(0) \Big(e_{\nu-1}(\pi+i2\tau\eta_1)-i\tau(e_{\nu-1})^2+\frac{i g_2\tau}{6}\Big)dx\nonumber&\\
&=2\lambda\lim_{\beta\to\infty}\int_A\beta(x)\bar{\partial}\bar{P}_\nu(0)\Big(\tau\Big(i 2\eta_1e_{\nu-1}-i(e_{\nu-1})^2+\frac{ig_2}{6}\Big)+\pi e_{\nu-1}\Big)dx\nonumber\\
&=0.
\end{align}
The last equality follows from the fact that the first term of the third line is zero for $\nu=2,3$ sectors, and the second term, which corresponds to the stress tensor, is suppressed by the factor $1/\beta$. In this way, the correction to the entanglement entropy vanishes in the low-temperature limit, which is consistent with the eq.(2.37) in\cite{Chen:2018eqk}.

\subsubsection{Bi-local modular Hamiltonian  }\label{eebd}
Unlike the local modular Hamiltonian, the form of the bi-local modular Hamiltonian is complicated. The correction of entanglement entropy from the bilocal modular Hamiltonian is difficult to study even using the resolvent. In this subsection, the low-temperature region $\beta\to\infty$ is evaluated. As the bi-local modular Hamiltonian for spin sector $\nu=3$ is zero in this limit \cite{Fries:2019ozf}, we focus on the $\nu=2$ sector at present.

In the low-temperature region $\beta\to\infty$, the bi-local modular Hamiltonian (\ref{localbi}) is simplified to \cite{Fries:2019ozf,Blanco:2019cet}
\begin{equation}\label{bbn}
\lim_{\beta\to\infty}\mathcal{K}_+^{bi-local}=-\int_{A_1}\int_{A_2}\frac{i}{\ell\sinh\frac{1}{2\ell}\log\frac{\Omega_A( x)}{\Omega_A( y)}}\psi^\dagger(x,\tilde{\tau})\psi(y,\tilde{\tau})dxdy.
\end{equation}
Using the perturbation method, the correction of the entanglement entropy from the bi-local modular Hamiltonian for $T\bar{T}$ deformed fermions is
\begin{equation}\label{t22}
{S_A^{bi-local}}^{(1)}=2\pi\lambda\Big(\langle\mathcal{T}_T\mathcal{K}_A^{bi-local}(\lambda)\rangle_\lambda-\langle\mathcal{T}_T\rangle_\lambda\langle\mathcal{K}_A^{bi-local}(\lambda)\rangle_\lambda\Big)\Big|_{\lambda=0}.
\end{equation}
 Applying the Wick's theorem, the first term of eq. (\ref{t22}) equals
\begin{align}\label{ttbarhb}
&\quad\langle\mathcal{T}_T\mathcal{K}_A^{bi-local}(\lambda)\rangle_0\nonumber\\
&=-\int_{A_1}\int_{A_2}\tilde{\beta}(x,y)\langle\psi^\dagger(x,\tilde{\tau})\psi(y,\tilde{\tau})\mathcal{T}_T\rangle dxdy\nonumber\\&=-\frac{i}{2}\int_{A_1}\int_{A_2}\tilde{\beta}(x,y)\int_{\mathcal{M}}\langle\psi^\dagger(x,\tilde{\tau})\psi(y,\tilde{\tau})T(z)\bar{T}(\bar{z})\rangle dxdyd^2z&\nonumber\\
&=-2\times\frac{i}{2}\int_{A_1}\int_{A_2}\int_\mathcal{M}\tilde{\beta}(x,y)\Big(\frac{\bar{\partial}\bar{P}_2(0)\partial P_2(0)P_2(x-y)}{2\pi i}-\frac{\bar{\partial}\bar{P}_2(0)\partial P_2(y-z)P_2(x-z)}{4\pi i}&\nonumber\\
&\quad+\frac{\bar{\partial}\bar{P}_2(0)\partial P_2(x-z)P_2(y-z)}{4\pi i}\Big)dxdyd^2z,&
\end{align}
 where an extra factor $2$ in the third line comes from the contributions from two different chiral fermion fields. Please refer to App. \ref{9.4} for detail calculation.
The eq. (\ref{ttbarhb}) turns out  to be
\begin{equation}\label{t2}
\lim_{\beta\to\infty}\langle\mathcal{T}_T\mathcal{K}_A^{bi-local}(\lambda)\rangle_0=-\frac{i}{\pi }\lim_{\beta\to\infty}\int_{A_1}\int_{A_2}\tilde{\beta}(x,y)(i\tau\bar{\partial}\bar{P}_2(0)\partial P_2(0)P_2(x-y)+\pi^3\bar{\partial}\bar{P}_2(0) P_2(x-y))dxdy.
\end{equation}
Taking ~(\ref{t2}) into ~(\ref{t22}), the leading order correction of entanglement entropy is
\begin{equation}\label{3.30}
\lim_{\beta\to\infty}{S_A^{bi-local}}^{(1)}(\lambda)=-2\lambda\pi^4\lim_{\beta\to\infty}\bar{\partial}\bar{P}_2(0)\langle K_A^{bi-local}(0)\rangle,
\end{equation}
where $\langle K_A^{bi-local}(0)\rangle=-\int\int_Adxdy\tilde{\beta}(x,y)\frac{P_2(x-y)}{\pi i}$ is the expectation value of undeformed bi-local modular Hamiltonian (\ref{bbn}). Therefore the leading order correction of entanglement entropy based on the bi-local modular Hamiltonian contributes a term which is the expectation value of the modular Hamiltonian of the undeformed fermion multiplied by a constant. From the analysis of the above two subsections, the leading order correction of entanglement entropy for $T\bar{T}$ deformed fermions on a torus from local(bilocal) modular Hamiltonian is proportional to the expectation value of local(bilocal) modular Hamiltonian for the undeformed fermions and its scale coefficient depends on the modular parameter $\tau$.

\subsection{A half-line on the plane }\label{e}
In the subsection, we turn to study the entanglement entropy of $T\bar{T}$ deformed CFT for the case where the chiral fermion is defined on an infinite line $\mathcal{L}$ at zero temperature, and the subsystem is chosen to be a half-line $P=\{x|x>0\}$. In such a setting,
the modular Hamiltonian is given by \cite{Bisognano:1975ih,Bisognano:1976za}
\begin{equation}
\mathcal{K}_P=-\int_P x T_{\tilde{\tau}\tilde{\tau}}(x)dx.
\end{equation}
Upto the second-order, the $T\bar{T}$ deformed action is \cite{He:2020cxp}
\begin{equation}\label{texpan}
\mathcal{I}_\lambda=\mathcal{I}_{CFT}-\frac{i\lambda}{2}\int_{\mathcal{L}} \Big(T^{(0)}(z)\bar{T}^{(0)}(\bar{z})-\frac{\lambda}{2}(T^{(0)}(z)\bar{T}^{(1)}(\bar{z})+T^{(1)}(z)\bar{T}^{(0)})(\bar{z})\Big)d^2z=\mathcal{I}_{CFT}-\lambda\mathcal{T}^{(0)}_T-\frac{\lambda^2}{2}\mathcal{T}^{(1)}_T,
\end{equation}
where $T^{(0)}$ and $T^{(1)}$ are the undeformed stress tensor and its leading order correction respectively. \footnote{We define $\mathcal{T}^{(0)}_T\equiv\frac{i}{2}\int_{\mathcal{L}} d^2z\Big(T^{(0)}(z)\bar{T}^{(0)}(\bar{z})\Big)$ and $\mathcal{T}^{(1)}_T\equiv-\frac{i\lambda}{4}\int_{\mathcal{L}} d^2z\Big((T^{(0)}(z)\bar{T}^{(1)}(\bar{z})+T^{(1)}(z)\bar{T}^{(0)})(\bar{z})\Big)$ in eq. (\ref{texpan}). and the deformed stress tensor is expanded as $T_{\mu\nu}^\lambda=\sum_{n=0}^\infty\frac{\lambda^n}{n!}T_{\mu\nu}^{(n)}$.}
From \cite{moosa}, the first order correction of the entanglement entropy is
\begin{equation}\label{f1}
\frac{1}{2\pi}\frac{dS_P(\lambda)}{d\lambda}=\langle \mathcal{T}^{(0)}_T\mathcal{K}_P(\lambda)\rangle_\lambda-\langle\mathcal{T}^{(0)}_T\rangle\langle\mathcal{K}_P(\lambda)\rangle_\lambda+\frac{\lambda}{2}(\langle \mathcal{T}^{(1)}_T\mathcal{K}_P(\lambda)\rangle_\lambda-\langle\mathcal{T}^{(1)}_T\rangle_\lambda\langle\mathcal{K}_P(\lambda)\rangle_\lambda).
\end{equation}
Similar to  the discussion in \cite{Rosenhaus:2014woa}, the leading order correction of the entanglement entropy vanishes since the integrand of $\langle \mathcal{T}^{(0)}\mathcal{K}_P\rangle_\lambda\propto(\langle\bar{T}\rangle+\langle T\rangle)$ which is zero in the plane. We now extend the calculation to the second-order correction, which contains the correction of the stress tensor for the deformed fermions.

Following the method in \cite{moosa}, the second-order derivative of the entanglement entropy behaves  as
\begin{align}\label{s2}
\frac{1}{2\pi}\frac{d^2S_P(\lambda)}{d\lambda^2}&=\langle\mathcal{T}^{(0)}_T\mathcal{T}^{(0)}_T\mathcal{K}_P(\lambda)\rangle_\lambda+\langle\mathcal{T}^{(0)}_T\frac{d\mathcal{K}_P(\lambda)}{d\lambda}\rangle_\lambda+\frac{1}{2}\langle\mathcal{T}^{(1)}_T\mathcal{K}_P(\lambda)\rangle_\lambda-\langle\mathcal{T}^{(0)}_T\mathcal{T}^{(0)}_T\rangle\langle\mathcal{K}_P(\lambda)\rangle_\lambda\nonumber&\\
&\quad-\langle\mathcal{T}^{(0)}_T\rangle\langle\mathcal{T}^{(0)}_T\mathcal{K}_P(\lambda)\rangle_\lambda-\frac{1}{2}\langle\mathcal{T}^{(1)}_T\rangle_\lambda\langle\mathcal{K}_P(\lambda)\rangle_\lambda+O(\lambda).
\end{align}
Plugging~(\ref{f1}) and~(\ref{s2}) into~(\ref{ET}), the entanglement entropy is
\begin{align}\label{s2o}
S_P(\lambda)&=S_P^{(0)}+\frac{\lambda^2}{2}\Big(\langle 3\pi\mathcal{T}^{(1)}_T\mathcal{K}_P(\lambda)\rangle_0+\langle2\pi\mathcal{T}^{(0)}_T\mathcal{T}^{(0)}_T\mathcal{K}_P(\lambda)\rangle_0+\langle2\pi\mathcal{T}^{(0)}_T\frac{d\mathcal{K}_P(\lambda)}{d\lambda}\rangle_0\Big)+O(\lambda^3),
\end{align}
where we used $\langle T(\omega)\bar{T}(\bar{z})\rangle_0=0=\langle T(\omega)\rangle_0$ in the plane.

The following three properties are needed to compute the second-order correction (\ref{s2o}).
The $T\bar{T}$ deformation is constructed from operators by point-splitting
\begin{equation}
 T(z)\bar{T}(\bar{z})=\lim_{\omega\to z}\Big( T(\omega)\bar{T}(z)-\Theta(\omega)\Theta(z)\Big).
\end{equation}
 Referring to \cite{McGough:2016lol}, the trace relation to the leading order in irrelevant parameter $\lambda$ is
\begin{equation}\label{tr}
\Theta^\lambda(z)=-\pi\lambda T^{(0)}\bar{T}^{(0)}+O(\lambda^2).
\end{equation}
In the complex plane, the conservation equation of energy-momentum tensor is \cite{Li:2020pwa}
\begin{align}\label{cem}
\partial_{\bar{z}}T_{zz}+\partial_zT_{z\bar{z}}=0,~~~~
\partial_{\bar{z}}T_{z\bar{z}}+\partial_zT_{\bar{z}\bar{z}}=0.
\end{align}
 Then the first term of the second-order correction in (\ref{s2o}) is,\footnote{Here we choose $v=x+i\tilde{\tau}$ and  $\bar{v}=x-i\tilde{\tau}$.}
\begin{align}\label{wht}
 \langle\mathcal{T}^{(1)}_T\mathcal{K}_P\rangle_0=\lim_{\omega_1\to z_1}\int_P\int_{\mathcal{R}^2}\langle 2\pi x T^{(0)}(\omega_1)\bar{T}^{(1)}(\bar{z}_1)(T^{(0)}(v)+\bar{T}^{(0)}(\bar{v}))\rangle_0 dxdz_1^2+h.c. ,
\end{align}
where the integrand turns out to be zero. To show this,  take the derivative of the integrand, using  (\ref{cem}), (\ref{tr}) and the point-splitting method,
\begin{align}\label{melon}
&\quad\partial_{z_1}\lim_{\omega_1\to z_1}\langle T^{(0)}(\omega_1)\bar{T}^{(1)}(\bar{z}_1)(T^{(0)}(v)+\bar{T}^{(0)}(\bar{v}))\rangle_0\nonumber+h.c.&\\
&=\pi\lambda\lim_{\omega_1\to z_1}\partial_{\bar{z}_1}\langle T^{(0)}(\omega)T^{(0)}(z_1)\bar{T}^{(0)}(\bar{z}_1)(T^{(0)}(v)+\bar{T}^{(0)}(\bar{v}))\rangle_0\nonumber+h.c.&\\
&=0
\end{align}
where we have used the conservation equation (\ref{cem}) and trace relation (\ref{tr}) in the second line of (\ref{melon}). The last line is obtained by factorization and $\langle T(z)\rangle=0$. Thus the integrand of  $\langle\mathcal{T}^{(1)}_T(\bar{z}_1)(T^{(0)}(v)+\bar{T}^{(0)}(\bar{v}))\rangle_0=C$, where $C$ is a anti-holomorphic function in $\bar{z}_1$ since $C$  can not depends on  $\bar{z}-\bar{v}$ from the Wick contraction. Then $C$ equals zero by the cluster decomposition principle. Thus we have $\langle\mathcal{T}^{(1)}_T\mathcal{K}_P\rangle_0=0$. \\
For the last term of (\ref{s2o}), it is \footnote{$T^{\lambda}$ represents the deformed stress tenor. In the plane, we have  $\frac{d\mathcal{K}_P(\lambda)}{d\lambda}=\frac{d\int_Px(T^\lambda(v)+\bar{T}^\lambda(\bar{v}))dx}{d\lambda}$ \cite{Rosenhaus:2014ula}. }
\begin{equation}\label{tflow}
\langle\mathcal{T}^{(0)}_T(\omega)\frac{d\mathcal{K}_P(\lambda)}{d\lambda}\rangle_0=\frac{d\langle\mathcal{T}^{(0)}_T(\omega)\mathcal{K}_P(\lambda)\rangle_0}{d\lambda}=\frac{d\langle \mathcal{T}^{(0)}_T(\omega)\int_Px(T^\lambda(v)+\bar{T}^\lambda(\bar{v}))dx\rangle_0}{d\lambda},
\end{equation}
where the derivative of the first term of the integrand in RHS of eq. (\ref{tflow}) is
\begin{align}
\frac{d\partial_{\bar{v}}\langle\mathcal{T}^{(0)}_T(\omega)T^\lambda(v)\rangle_0}{d\lambda}&=\frac{d\partial_{v}\langle\mathcal{T}^{(0)}_T(\omega)\Theta^\lambda(v)\rangle_0}{d\lambda}=\frac{-\pi d(\lambda\partial_{v}\langle\mathcal{T}^{(0)}_T(\omega)\mathcal{T}^{(0)}_T(v)\rangle_0)}{d\lambda}=\frac{\pi c^2}{(\omega-v)^5(\bar{\omega}-\bar{v})^4}
\end{align}
Taking a similar analysis as $\langle\mathcal{T}^{(1)}_T\mathcal{K}_P\rangle_0=0$, we can obtain $\frac{d\langle\mathcal{T}^{(0)}_T(\omega)T^\lambda(v)\rangle_0}{d\lambda}=\frac{\pi c^2}{3(\omega-v)^5(\bar{\omega}-\bar{v})^3}$ which is consistent with eq. (4.9) in \cite{Kraus:2018xrn}. Meanwhile we can  acquire $\frac{d\langle\mathcal{T}^{(0)}_T(\omega)\bar{T}^\lambda(\bar{v})\rangle_0}{d\lambda}=\frac{\pi c^2}{3(\omega-v)^3(\bar{\omega}-\bar{v})^5}$. Thus (\ref{tflow}) is simplified to be
\begin{equation}\label{wuyu}
\langle\mathcal{T}^{(0)}_T(\omega)\frac{d\mathcal{K}_P(\lambda)}{d\lambda}\rangle_0=\int_\mathcal{L}\int_P\pi c^2x\Big(\frac{1}{3(\omega-v)^5(\bar{\omega}-\bar{v})^3}+\frac{1}{3(\omega-v)^3(\bar{\omega}-\bar{v})^5}\Big)dxd^2\omega.
\end{equation}
From the calculation of (\ref{wuyu}) presented in  App. \ref{c.3}, we obtain
\begin{equation}\label{wlgq}
\langle\mathcal{T}^{(0)}_T(\omega)\frac{d\mathcal{K}_P(\lambda)}{d\lambda}\rangle_0=0.
\end{equation}
Following the same way in deriving $\langle\mathcal{T}^{(1)}_T\mathcal{K}_P\rangle_0=0$ and eq. (\ref{wlgq}), one can obtain
\begin{equation}\label{ga}
\langle\mathcal{T}^{(0)}_T\mathcal{T}^{(0)}_T\mathcal{K}_P\rangle_0=0,
\end{equation}
where the calculation is presented in App. \ref{c.3}.
Now substituting $\langle\mathcal{T}^{(1)}_T\mathcal{K}_P\rangle_0=0$, (\ref{ga}) and (\ref{wlgq}) into~(\ref{s2o}), the second-order correction to the entanglement entropy of the half-line is zero.
To sum up, we find that the leading order and second-order correction of the entanglement entropy for the $T\bar{T}$ deformed fermions on a half-line vanish.

\section{Entanglement entropy of a local bilinear operator deformed fermions}
It is an interesting question to study the time evolution of the entanglement entropy under injection of local excitation in free fermion. {It helps us to understand the time evolution mechanism of the system. As an example, the authors of \cite{Calabrese:2007rg, Calabrese:2005in} suggested that the propagation of the quasiparticles can be used to describe the time evolution of entanglement entropy.} {We would like to figure out the entanglement entropy of a single interval on a torus for free fermion under the deformation of a local bilinear operator by adding a source localized in space and time in Sec. \ref{localdf}. In Sec. \ref{f} under the same deformation, the time evolution of entanglement entropy for a single interval is calculated in a system with a time-dependent boundary, i.e., moving mirror setting \cite{Davies:1976hi}\cite{Reyes:2021npy}.} It gives us some insights into hawking radiation \cite{Davies:1976hi, Hawking:1974rv, Hawking:1975vcx} from the field theory side. One can refer to recent progresses \cite{Akal:2020twv, Reyes:2021npy, Akal:2021foz, Bianchi:2022ulu} in this direction.
\subsection{Single interval on a torus }\label{localdf}
In this section, we analyze the entanglement entropy to a single interval $A=(a,b)$ in case of local deformation for the chiral fermion on a torus.
We perturb the system with a local term, i.e. adding a source supported at a certain point inside the interval $[a,b]$
\begin{equation}
\mathcal{I}'_{\psi}=\mathcal{I}_{CFT}-\lambda\int_{\mathcal{M}}\delta^{(2)}(x-z_0)\psi^\dagger(w,\tilde{\tau})\psi(w,\tilde{\tau})d^2x=\mathcal{I}_{CFT}-\lambda \psi^\dagger(w_0,\tilde{\tau}_0)\psi(w_0,\tilde{\tau}_0),
\end{equation}
where $z_0=w_0+i\tilde{\tau},w_0\in[a,b]$. We have $\mathcal{T}_L=\psi^\dagger(w_0,\tilde{\tau}_0)\psi(w_0,\tilde{\tau}_0)$.

According to the leading order correction of entanglement entropy (\ref{dh}) and the local modular Hamiltonian (\ref{localh}), it yields
\begin{align}\label{mld}
\langle\psi^\dagger(w_0,\tilde{\tau}_0)\psi(w_0,\tilde{\tau}_0)T^{local}_A(x)\rangle_0=\langle \psi^\dagger(w_0,\tilde{\tau}_0)\psi(w_0,\tilde{\tau}_0)\rangle_0\langle T^{local}_A(x)\rangle_0,
\end{align}
where the other terms produced by contraction of the stress tensor with $\psi^\dagger(w_0,\tilde{\tau}),\psi(w_0,\tilde{\tau})$ are cancelled out in (\ref{mld}).
Taking (\ref{mld}) into (\ref{dh}), the leading order correction of entanglement entropy based on the local modular Hamiltonian is zero.

 Bringing~(\ref{localbi}) into~(\ref{t22}) under local operator deformation, the leading order correction of entanglement entropy from the bi-local modular Hamiltonian becomes
 \begin{align}\label{rem}
& {S_A^{bi-local}}^{(1)}(\lambda)\nonumber\\
=&\frac{\lambda}{\pi}\int_A\sum_{k\in\mathbb{Z} \backslash \{0\}}(\pm 1)^k\tilde{\beta}(x,x_k(x))(P_\nu(x-w_0)P_\nu(x_k(x)-w_0)\delta\Big(x-x_k+\beta \frac{1}{2\pi(b-a)}\log\frac{\Omega_A(x|\tau)}{\Omega_A(x_k|\tau)}-k\Big)dx.
 \end{align}
{In terms of (\ref{rem}), the correction of entanglement entropy based on the bi-local modular Hamiltonian depends on the correlation function of the local bilinear operators.
 Since the correlation function represented by $P_\nu(x)$ is a periodic function,} the leading order correction of entanglement entropy based on the bi-local modular Hamiltonian is periodic concerning the location of the inserting operator.

 In the low-temperature region $\beta\to\infty$, as discussed in~(\ref{eebd}), the correction of the entanglement entropy for spin sector $\nu=2$ yields
 \begin{align}\label{haa}
  \lim_{\beta\to\infty}{S_A^{bi-local}}^{(1)}(\lambda)&=\frac{\lambda}{\pi} \int_{A_1}\int_{A_2}\frac{i\beta\pi}{\ell\sinh\frac{1}{2\ell}\log\frac{\Omega_A( x_1)}{\Omega_A( y_2)}}P_\nu(x_1-w_0)P_\nu(y_2-w_0)dx_1dy_2&\nonumber\\
 &=0.
\end{align}
{This final step of eq. (\ref{haa}) follows from the fact that the above equation is anti-symmetry with respect to $x_1$ and $y_2$. }


\subsection{ Single interval in the Moving mirror }\label{f}
In this subsection, we would like to calculate the entanglement entropy of chiral fermion in a single interval $A=(a,b)$ with a time-dependent boundary. In particular, we study chiral free fermion with a moving mirror boundary under local bilinear operator deformation \cite{Davies:1976hi}.

 Consider the action of massless Dirac spinors $\psi_\pm$ over a region $\mathcal{M}$ of $1+1$-dimensional Minkowski spacetime with a boundary $\partial\mathcal{M}$ defined by $v=g(u)$.  In light-cone coordinate $u=x_+=t-x,v=x_-=t+x$, the action reads \cite{Reyes:2021npy}
\begin{equation}
\mathcal{I}_{mirror}=\frac{i}{2}\int_{\mathcal{M}}(\psi^\dagger_-\partial_+\psi_-+\psi^\dagger_+\partial_-\psi_+)dxdt.
\end{equation}
Upon variation, the action gives a total derivative term along the boundary that
\begin{equation}
\mathcal{I}_B=\frac{i}{2}\int_{\partial \mathcal{M}}(\psi^\dagger_-\delta \psi_--g'(x^-)\psi_+^\dagger\delta\psi_+)dx^-+h.c.
\end{equation}
Here we study the system constrained to the correct part of the boundary. Therefore the right and left movers $\psi(u),\psi(v)$ are regarded as the outgoing and incoming modes.
 The total derivative term $\mathcal{I}_B$ of the action vanishes when $\partial{M}$ acts as a ``mirror'' by imposing the reflected boundary condition \cite{Reyes:2021npy}, which is
\begin{equation}\label{aha}
\psi_+(u)=-\sqrt{g'(u)}\psi_-(g(u)).
\end{equation}
Following \cite{WanMokhtar:2018lwi} the in-mode $\psi_k^{in}$ ansatz whose left moving part is proportional to $e^{-ikv}U_2$,  the reflected boundary condition (\ref{aha}) implies that
\begin{equation}\label{slsr}
\psi_k^{in}=-\sqrt{\frac{g'(u)}{2\pi}}e^{-i k g(u)}U_1+\frac{1}{\sqrt{2\pi}}e^{-ik v}U_2,
\end{equation}
where the orthogonal spinor basis is $\{ U_1=\begin{pmatrix}1\\0\end{pmatrix},U_2=\begin{pmatrix}0\\1\end{pmatrix}\}$.\\
The chiral fermion can then be expanded as
\begin{equation}\label{qslsr}
\psi(u,v)=\int^\infty_0(a^{in}_k\psi^{in}_k (u,v)+b^{in\dagger}_{k}\psi^{in}_{-k} (u,v))dk,
\end{equation}
here $a^{in}_k,b^{in}_k$ and $a^{in\dagger}_{k},b^{in\dagger}_{k}$ are the annihilation and creation operators satisfying
\begin{equation}
\{a^{in}_k,a^{in\dagger}_{k'}\}=\{b^{in}_k,b^{in\dagger}_{k'}\}=\delta(k-k'), \quad\text{for} \quad k,k'>0.
\end{equation}
These operators define a normalised in-vacuum state $|0_{in}\rangle$ which means
\begin{equation}
    a^{in}_k|0_{in}\rangle= b^{in}_k|0_{in}\rangle=0, \quad\text{for} \quad k>0.
\end{equation}
Taking~(\ref{slsr}) and~(\ref{qslsr}) into~(\ref{St12}), the renormalized stress tensor of chiral fermion becomes
\begin{equation}\label{stss}
\langle T_{uu}(u)\rangle\propto \Big(\frac{g'''(u)}{g'(u)}-\frac{3}{2}(\frac{g''(u)}{g'(u)})^2\Big),
\end{equation}
which is the Schwarzian derivative of $g(u)$ with respect to $u$.

In the moving mirror setting, the action induced by a local bilinear operator deformation is
\begin{equation}\label{toil}
\tilde{\mathcal{I}}_{\psi}=\mathcal{I}_{mirror}-\frac{i\lambda}{2}\int_{\mathcal{M}}\delta^{(2)}(z-z_0)\psi^\dagger(x_0,t_0)\psi(x_0,t_0)d^2z=\mathcal{I}_{mirror}-\lambda \psi^\dagger(x_0,t_0)\psi(x_0,t_0),
\end{equation}
where $x_0$ is the insertion point for the local operator and we define $\mathcal{T}_r=\psi^\dagger(x_0,t_0)\psi(x_0,t_0)$.\\
According to the perturbation method presented in Sec.~\ref{b}, the correction of entanglement entropy is expanded as
\begin{align}\label{jcs}
S_A(\lambda)&=S_A^{(0)}+\lambda(\langle2\pi\mathcal{T}_r\mathcal{K}_A(\lambda)\rangle_0-\langle\mathcal{T}_r\rangle_0\langle2\pi\mathcal{K}_A(\lambda)\rangle_0)+\lambda^2(\langle\pi\mathcal{T}_r\mathcal{T}_r\mathcal{K}_A(\lambda)\rangle_0\nonumber&\\&\quad+\langle\pi\mathcal{T}_r\frac{d\mathcal{K}_A(\lambda)}{d\lambda}\rangle_0-\langle\pi\mathcal{T}_r\mathcal{T}_r\rangle_0\langle\mathcal{K}_A(\lambda)\rangle_0-\langle\mathcal{T}_r\rangle_0\langle\pi\mathcal{T}_r\mathcal{K}_A(\lambda)\rangle_0\Big)+O(\lambda^3)
\end{align}
The equation implies that the correction of entanglement entropy relies on the local bilinear operator's correlation function and the modular Hamiltonian derivative.

From the eq. (\ref{ker}), the modular Hamiltonian $\mathcal{K}_A$ takes the form
\begin{equation}\label{kerh}
\mathcal{K}_A=\int_{\mathcal{M}} \psi^\dagger(x)k(x,y)\psi(y)dxdy.
\end{equation}
where the component of kernel in the presence of moving mirror is \cite{Reyes:2021npy}
\begin{equation}\label{kernel}
k_{ij}(x,y)=-2\pi\delta(Z(q_i(x))-Z(q_j(x)))G_{ij}(x,y),
\end{equation}
where $G_{ij}=\langle\psi_i\psi_j^\dagger\rangle$ with $i,j=\pm$ and $q_+(x)=g(x_+),q_-(x)=x_-$. Function $Z(x)=\frac{1}{2\pi}\log{(-\frac{G(x,b_-)}{G(x,a_-)}\frac{G(x,g(a_+))}{G(x,g(b_+))}})$.
Since different chiral fermions are entangled, the kernel is not diagonal \cite{Reyes:2021npy}.

As the mirror trajectory generally breaks the system's conformal symmetry, the moving mirror setup is not the traditional boundary conformal field theory (BCFT). Instead of studying the moving mirror by transforming the moving trajectory into static BCFT in \cite{Akal:2021foz}, we analyze the correlation function in the moving mirror setup through the Wick contraction.
From the eq. (\ref{kernel}), the modular Hamiltonian depends on the trajectory of the mirror. We investigate the uniformly accelerated mirror following the trajectory $t^2-x^2=-R^2$, which corresponds to
\begin{equation}\label{mirror}
g(u)=-\frac{R^2}{u}.
\end{equation}
Substituting (\ref{mirror}), (\ref{qslsr}) into (\ref{slsr}), the two-point correlation functions of chiral fermion are given by
\begin{align}\label{aong}
\langle\psi_-(u_1)\psi^*_-(u_2)\rangle_0&=\int_0^\infty dk\frac{\sqrt{g'(u_1)g'(u_2)}}{2\pi}e^{-ikg(u_1)}e^{ikg(u_2)}\nonumber&\\
&=-\frac{i\sqrt{g'(u_1)g'(u_2)}}{2\pi(g(u_1)-g(u_2))}\nonumber&\\
&=\frac{i}{2\pi(u_2-u_1)},
\end{align}
and
{\begin{align}\label{awang}
\langle\psi_+(u_1)\psi^*_+(u_2)\rangle_0&=\int_0^\infty dk\frac{1}{2\pi}e^{-ikv_1}e^{ikv_2}\nonumber&\\
&=-\frac{i}{2\pi(v_1-v_2)}.
\end{align}}
In the limit $t\to\infty$, it shows that the correlation functions become
\begin{equation}\label{pm}
G_{\pm\mp}\to 0.
\end{equation}
Thus the off-diagonal elements of eq. (\ref{kernel}) vanish.
Plugging (\ref{aong}), (\ref{awang}) and (\ref{kernel}) into (\ref{kerh}), the undeformed modular Hamiltonian $\mathcal{K}^{(0)}_A$ can be written in a compact form as
\begin{equation}\label{mmhe}
    \mathcal{K}^{(0)}_A={\mathcal{K}}_{++}^{(0)}+{\mathcal{K}}_{--}^{(0)},
\end{equation}
that
\begin{equation}\label{mmm}
{\mathcal{K}}_{++}^{(0)}=\int_\mathcal{M}\frac{2\pi}{Z'(g(u))}T_{uu}du,\quad {\mathcal{K}}_{--}^{(0)}=\int_\mathcal{M}\frac{2\pi}{Z'(v)}T_{vv}dv,
\end{equation}
where $Z'(g(u))$ or $Z'(v)$ represents the derivative of function $Z$ with respect to $u$ or $v$. Comparing (\ref{mmm}) with~(\ref{localh}), the effective temperature $\beta(u)=\frac{2\pi}{Z'(g(u))}$ depends on the trajectory of the moving mirror.
Plugging (\ref{aong}) and (\ref{awang}) into (\ref{toil}), we obtain
\begin{equation}\label{gaga}
    \langle\mathcal{T}_r\rangle_0\propto(\langle\psi_-(w_1)\psi^*_-(w_2)\rangle_0+\langle\psi_+(w_1)\psi^*_+(w_2)\rangle_0),
\end{equation}
with
\begin{align}\label{wong}
\langle\psi_-(w_1)\psi^*_-(w_2)\rangle_0&=\int_0^\infty dk\frac{\sqrt{g'(w_1)g'(w_2)}}{2\pi}e^{-ikg(w_1)}e^{ikg(w_2)}\nonumber&\\
&=-\frac{i\sqrt{g'(w_1)g'(w_2)}}{2\pi(g(w_1)-g(w_2))}\nonumber&\\
&=-{\frac{1}{4\pi\epsilon},}
\end{align}
and
\begin{equation}\label{wang}
\langle\psi_+(w_1)\psi^*_+(w_2)\rangle_0=-\frac{1}{4\pi\epsilon},
\end{equation}
here we define $w_1=t_0-x_0+i\epsilon,w_2=t_0-x_0-i\epsilon$ and apply the regularization to remove the divergent pieces presented in the last line of eq. (\ref{wong}) and eq. (\ref{wang}).
By plugging (\ref{mmhe}), (\ref{toil}) and (\ref{gaga}) into (\ref{jcs}), one can show the correction of the entanglement entropy (\ref{jcs}) is
\begin{equation}\label{mirrorh}
\lim_{t\to\infty}\Delta S_A(\lambda)=\lambda^2\langle\pi\mathcal{T}_r\frac{d\mathcal{K}_A(\lambda)}{d\lambda}\rangle_0,
\end{equation}
where the other terms in (\ref{jcs}) are simply zero upon the Wick contraction. Therefore (\ref{mirrorh}) implicates that the second-order correction of entanglement entropy depends on contraction of local bilinear operator and the deformed modular Hamiltonian.

Under the local bilinear operator deformation, we can calculate (\ref{mirrorh}) by comparing the one-point function $\langle\mathcal{T}_r\rangle_\lambda$ in operator formalism with this one-point function in path integral formalism.\\
In the path integral formalism, the one-point function  $\langle\mathcal{T}_r\rangle_\lambda$ is expanded as
\begin{equation}\label{ant}
    \langle\mathcal{T}_r\rangle_\lambda= \langle\mathcal{T}_r\rangle_0+\lambda\langle\mathcal{T}_r\mathcal{T}_r\rangle_0-\lambda\langle\mathcal{T}_r\rangle_0\langle\mathcal{T}_r\rangle_0+O(\lambda^2).
\end{equation}
While the one-point function in operator formalism, it shows that
\begin{align}\label{anm}
     \langle\mathcal{T}_r\rangle_\lambda&=\text{Tr}_{A\cup\bar{A}}\Big(\rho(\lambda)\mathcal{T}_r\Big)=\text{Tr}_{A}  \Big(\rho_A(\lambda)\mathcal{T}_r\Big)=\text{Tr}_{A} \Big(\frac{e^{-2\pi {\mathcal{K}_A}^{(0)}-2\pi\lambda\frac{d\mathcal{K}_A(\lambda)}{d\lambda}\big|_{\lambda=0}+O(\lambda^2)}}{Z_\lambda}\mathcal{T}_r\Big)\nonumber\\
     &=\langle\mathcal{T}_r\rangle_0-2\pi\lambda\text{Tr}_{A} \Big(\frac{e^{-2\pi {\mathcal{K}_A}^{(0)}}}{Z_0}\frac{d\mathcal{K}_A(\lambda)}{d\lambda}\big|_{\lambda=0}\mathcal{T}_r\Big)-\lambda\langle\mathcal{T}_r\rangle_0\langle\mathcal{T}_r\rangle_0\nonumber\\
     &=\langle\mathcal{T}_r\rangle_0-2\pi\lambda\text{Tr}_{A} \Big(\frac{e^{-2\pi {\mathcal{K}_A}^{(0)}}}{Z_0}\mathcal{T}_r\frac{d\mathcal{K}_A(\lambda)}{d\lambda}\big|_{\lambda=0}\Big)-\lambda\langle\mathcal{T}_r\rangle_0\langle\mathcal{T}_r\rangle_0\nonumber\\
     &=\langle\mathcal{T}_r\rangle_0-2\pi\lambda\langle\mathcal{T}_r\frac{d\mathcal{K}_A(\lambda)}{d\lambda}\rangle_0-\lambda\langle\mathcal{T}_r\rangle_0\langle\mathcal{T}_r\rangle_0.
\end{align}
In deriving the second and third line of the equation, we have used the cyclic permutation property of a trace and take only terms which are functions of the undeformed modular Hamiltonian $\mathcal{K}_A^{(0)}$ with only a single operator $\frac{d\mathcal{K}_A(\lambda)}{d\lambda}\big|_{\lambda=0}$.
\footnote{To linear order in small perturbation of $\langle\mathcal{T}_r\rangle_\lambda$, we have to prove
\begin{equation}\label{flee}\text{Tr}_A\Big(e^{-2\pi\mathcal{K}_A^{(0)}-2\pi\lambda\frac{d\mathcal{K}_A(\lambda)}{d\lambda}\big|_{\lambda=0}}\mathcal{T}_r\Big)=\text{Tr}_A\Big(e^{-2\pi\mathcal{K}_A^{(0)}}e^{-2\pi\lambda\frac{d\mathcal{K}_A(\lambda)}{d\lambda}\big|_{\lambda=0}}\mathcal{T}_r\Big)
\end{equation}
. According to the BCH formula $e^Ae^B=e^{A+B+\frac{1}{2}[A,B]+\frac{1}{12}[A,[A,B]]+\frac{1}{12}[B,[B,A]]...}$, it is equivalent to prove
\begin{equation}\label{uu}
\text{Tr}_A\Big({ \mathcal{K}_A^{(0)}}^m[\mathcal{K}_A^{(0)},\frac{d\mathcal{K}_A(\lambda)}{d\lambda}\big|_{\lambda=0}]{\mathcal{K}_A^{(0)}}^n\mathcal{T}_r\Big)=0
\end{equation}
which is the general term in (\ref{flee}). We can obtain (\ref{uu}) from the cyclic property of the operator under the trace and $\lim_{t\to\infty}[\mathcal{K}_A^{(0)},\mathcal{T}_r]=0$  from (\ref{kerh}). }
Then comparing eq. (\ref{anm}) with eq. (\ref{ant}), we obtain
\begin{equation}\label{err}
    \langle\mathcal{T}_r\frac{d\mathcal{K}_A(\lambda)}{d\lambda}\rangle_0=-\frac{1}{2\pi}\langle\mathcal{T}_r\mathcal{T}_r\rangle_0.
\end{equation}
Taking (\ref{gaga}),(\ref{wang}), (\ref{wong}) and (\ref{err}) into (\ref{mirrorh}), the correction of entanglement entropy vanishes as
\begin{equation}
\lim_{t\to\infty}\Delta S_A(\lambda)\propto\langle\mathcal{T}_r\mathcal{T}_r\rangle_0=0.
\end{equation}

 According to (\ref{mirrorh}), the correction of the entanglement entropy of chiral free fermions under the local bilinear operator deformation on a torus depends on the expectation value of the deformed stress tensor. In the late time limit, the correction vanishes since there is no radiation of stress tensor in the uniformly accelerated mirror at $t\to\infty$. That is to say, the deformed system achieves equilibrium in late time. A similar situation has been discussed in \cite{Davies:1976hi}, which is associated with the scalar field in the moving mirror setting. One can refer to a more recent discussion \cite{Reyes:2021npy} about the entanglement entropy of chiral fermion in the radiation process.


\section{Entanglement entropy of mass deformed fermions}\label{mee}
In the section, we will study the entanglement entropy  of a single interval $A=(a,b)$ on a torus in mass deformed fermion theory. In particular, we apply the perturbative field theory approach to calculate the entanglement entropy of subsystem in the deformed theory. The deformed action reads
\begin{equation}
\mathcal{I}_{\psi}=\mathcal{I}_{CFT}-m\int_{\mathcal{M}}\bar{\psi}(x,\tilde{\tau})\psi(x,\tilde{\tau})d^2x.
\end{equation}
Comparing this deformed action with~(\ref{actiond}), we have $\mathcal{T}_m=\int_{\mathcal{M}}\bar{\psi}(x,\tilde{\tau})\psi(x,\tilde{\tau})d^2x$ and $\lambda=m$ at present case. Expanding the entanglement entropy to the second-order with respect to the mass parameter, one obtains
\begin{align}\label{eem}
S_A(m)&={S_A}_{(0)}+m(\langle2\pi\mathcal{T}_m\mathcal{K}_A(m)\rangle_0-\langle\mathcal{T}_m\rangle_0\langle2\pi\mathcal{K}_A(m)\rangle_0)+m^2(\langle\pi\mathcal{T}_m\mathcal{T}_m\mathcal{K}_A(m)\rangle_0\nonumber&\\&\quad+\langle\pi\mathcal{T}_m\frac{d\mathcal{K}_A(m)}{dm}\rangle_0-\langle\pi\mathcal{T}_m\mathcal{T}_m\rangle_0\langle\mathcal{K}_A(m)\rangle_0-\langle\mathcal{T}_m\rangle_0\langle\pi\mathcal{T}_m\mathcal{K}_A(m)\rangle_0\Big)+O(m^3).
\end{align}
The first-order correction of entanglement entropy is zero as $\langle T\mathcal{T}_m\rangle_0=0$ and $\langle\mathcal{T}_m\rangle_0=0$ in terms of the contraction of chiral fields. The same result can be also obtained from the eq. (57) in \cite{Herzog:2013py}.

In the following, we study the integrals in second order correction of entanglement entropy from the local modular Hamiltonian (\ref{hmer}).
The integrand in the first term of second-order correction in (\ref{eem}) is proportional to $\langle\mathcal{T}_m\mathcal{T}_m(T+\bar{T})\rangle_0$,
 \begin{align}\label{stran}
 \langle\mathcal{T}_m\mathcal{T}_mT(z)\rangle_0&=2\int_{\mathcal{M}_1}\int_{\mathcal{M}_2}\Big(\frac{\partial P_\nu(z-z_2)P_\nu(z-z_1)\bar{P}_\nu(\bar{z}_1-\bar{z}_2)}{8\pi^2}-\frac{\partial P_\nu(z-z_1)P_\nu(z-z_2)\bar{P}_\nu(\bar{z}_1-\bar{z}_2)}{8\pi^2}\nonumber&\\
 &\quad-\frac{\partial P_\nu(0)P_\nu(z_1-z_2)\bar{P}_\nu(\bar{z}_1-\bar{z}_2)}{4\pi^2}\Big)dx_1^2dx_2^2\nonumber&\\
 &=\int_{\mathcal{M}_1}\int_{\mathcal{M}_2}\Big(\frac{\partial P_\nu(z-z_2)P_\nu(z-z_1)\bar{P}_\nu(\bar{z}_1-\bar{z}_2)}{2\pi^2}\Big)dx_1^2dx_2^2+\langle T(z)\rangle_0\langle\mathcal{T}_m\mathcal{T}_m\rangle_0.
\end{align}
 For the fermions in $\nu=2$ sector at the low temperature limit $\beta\to\infty$, one can obtain that
\begin{equation}\label{weyl}
   \lim_{\beta\to\infty} \langle\mathcal{T}_m\mathcal{T}_m\rangle_0=2\lim_{\beta\to\infty}\int_{\mathcal{M}_1}\int_{\mathcal{M}_2}(P_2(z_1-z_2)\bar{P}_2(\bar{z}_1-\bar{z}_2))dx_1^2dx_2^2\sim 2\beta\pi\ln\pi,
\end{equation}
where $``\sim"$ means that the result is regularized by dropping the divergent terms.
And
\begin{equation}\label{codozz}
     \lim_{\beta\to\infty}\int_{\mathcal{M}_1}\int_{\mathcal{M}_2}\Big(\frac{\partial P_2(z-z_2)P_2(z-z_1)\bar{P}_2(\bar{z}_1-\bar{z}_2)}{2\pi^2}\Big)dx_1^2dx_2^2=\frac{\pi^4\ln\pi}{2}
\end{equation}
where the Wick contraction and two-point function for fermion fields \cite{DiFrancesco:1997nk} are used. Please refer to App. \ref{9.5} for details of the calculation.

Substituting (\ref{stran}), (\ref{weyl}) and (\ref{codozz}) into (\ref{eem}), the undetermined term in the second-order correction of the entanglement entropy is $\langle\pi\mathcal{T}_m\frac{d\mathcal{K}^{local}_A(m)}{dm}\rangle_0$. It is quite hard to obtain the derivative of the modular Hamiltonian in $\langle\pi\mathcal{T}_m\frac{d\mathcal{K}^{local}_A(m)}{dm}\rangle_0$ on a torus.

In summary, we perturbatively compute the entanglement entropy of a single interval in mass deformed free fermion. The leading order correction of entanglement entropy for mass deformed fermions vanishes. To calculate the second-order correction of the entanglement entropy, it is necessary to fix the derivative of the modular Hamiltonian with respect to mass. It is quite a hard attempt to do so. Hopefully, we can figure out a concrete answer in the future.

\section{Conclusions and perspectives}\label{g}
In this work,  we calculate the modular Hamiltonian of free fermion with three kinds of deformation, including $T\bar{T}$ deformation, mass deformation, and local bilinear operator deformation. It shows that the modular Hamiltonian is an effective tool to analyze entanglement entropy under the perturbation framework, even for the system on a torus.

The modular Hamiltonian and entanglement entropy in free fermion theory with $T\bar{T}$ deformation are perturbatively evaluated in Sec. \ref{eottt} and Sec. \ref{e}. In Sec. \ref{eottt}, we find that there exists a leading order correction of entanglement entropy on a torus. The leading order correction to entanglement entropy of $T\bar{T}$ deformed fermions always vanishes once the entangling surface is a plane or a sphere in a flat space. The associating local (bi-local) modular Hamiltonian of $T\bar{T}$ deformed fermions on a torus acquires a new contribution which is proportional to the expectation value of local(bi-local) modular Hamiltonian for the undeformed fermions from the eq. (\ref{ttbarh}) and (\ref{ttbarhb}). The scale factor in the deformed modular Hamiltonian depends on the modular parameter of a torus. As a consistency check, one can take the cylinder limit and reduce it to the leading order correction of entanglement entropy obtained by replica trick \cite{Chen:2018eqk}. In Sec. \ref{e}, we also demonstrate that the second-order correction of the entanglement entropy of a half-plane in free fermion with a $T\bar{T}$ deformation vanishes.

The modular Hamiltonian in the free fermion field theory deformed by a local bilinear operator is investigated in  Sec. \ref{localdf}. The leading order corrections to the entanglement entropy from the local and bi-local modular Hamiltonian are vanishing at the low-temperature limit. The modular Hamiltonian in the deformed free fermion has been evaluated in the moving mirror setting. It shows a time-dependent behavior due to the satisfying mirror boundary condition. In particular, since there is no radiation in the time-dependent vacuum \cite{Davies:1976hi}, the deformed entanglement entropy will vanish within the late time limit.

As for mass deformation in free fermion field theory on a torus, the correction of entanglement entropy is evaluated in Sec. \ref{mee}. There is no leading order correction to the entanglement entropy of mass deformed fermions. A similar observation was obtained in  \cite{Herzog:2013py}, where the Sine-Gordon Model acquires no leading order correction of entanglement entropy on a torus. For the second-order correction of the entanglement entropy, we left an undetermined term $\langle\pi\mathcal{T}_m\frac{d\mathcal{K}^{local}_A(m)}{dm}\rangle_0$ for the local Hamiltonian part. One may need to study the correlation function under mass deformed on a torus to analyze the entanglement entropy like the planar case \cite{Casini:2009vk}.

This work evaluates the modular Hamiltonian and entanglement entropy in free fermion field theory with the three typical deformations. Fully understanding the structure of modular Hamiltonian in generic quantum field theories is quite challenging because there is no good non-perturbative approach to evaluate, even in free field theory. It is an interesting problem to study local quench triggered by a local operator in CFTs with time-dependent boundary conditions to mimic the Hawking radiation \cite{Akal:2020twv, Reyes:2021npy, Akal:2021foz}. One can calculate the time evolution of modular Hamiltonian or entanglement entropy to extract the characteristic dynamical behavior of radiation pairs.

\section*{Acknowledgments}
We would like to thank Miao He, I.~A.~Reyes, Hong-An Zeng, Yu-Xuan Zhang, Taishi Kawamoto for valuable discussions related to this work. S.H. also would like to appreciate the financial support from Jilin University and Max Planck Partner group, as well as the Natural Science Foundation of China Grants No.~12075101, No.~12047569.  Y.S. is supported by the National Natural Science Foundation of China Grants No. 12105113.

\appendix
\section{Modular Hamiltonian on a cylinder}
\subsection*{Zero temperature}\label{9.1}
The Modular Hamiltonian for a single interval on a cylinder at zero temperature is
\begin{equation}
\mathcal{K}_V=\int_C\frac{T_{\tilde{\tau}\tilde{\tau}}(x)}{f'(x)}dx+\frac{c}{12\pi}\int_C\frac{\{f(x);x\}}{f'(x)}dx
\end{equation}

For the second term of the above equation, we have that
\begin{align}
&\int_C\frac{c}{12\pi}\{f(x);x\}(\frac{df(x)}{dx})^{-1}dx\nonumber\\
=&\frac{c}{12\pi}[\frac{1}{2(-1+e^{4i\pi R/L})L}\Big(-L+e^{4i\pi R/L}L-2(1+e^{4i\pi R/L})i\pi R&\nonumber\\
&-(-1+e^{4i\pi R/L})L\log(\frac{-2i\pi\epsilon}{L})+(-1+e^{4i\pi R/L})L\log(2\sinh(\frac{2i\pi R}{L}))\Big)&\nonumber\\
&-\frac{1}{2(-1+e^{4i\pi R/L})L}\Big(-e^{4i\pi R/L}L+L+2(1+e^{4i\pi R/L})i\pi R&\nonumber\\
&-(-1+e^{4i\pi R/L})L\log(1-e^{4i\pi R/L})+(-1+e^{4i\pi R/L})L\log(e^{2i\pi R/L}\frac{2\pi\epsilon}{L})\Big)]&\nonumber\\
=&\frac{c}{12\pi}\log(\frac{L}{\pi\epsilon}\sin\frac{2\pi R}{L})+\frac{c}{12\pi}-\frac{c}{12\pi}\frac{2\cos(2\pi R/L)i\pi R}{\sin(2\pi R/L)L}-\frac{i\pi c}{12},&
\end{align}
where the last two terms of the above equation is a imaginary number. We take the real part of the above equation.

Taking  ~(\ref{stress}) and $f(x)$ defined in the context into the first term of the cylinder's modular Hamiltonian, the result is
\begin{align}
\int_C\frac{T_{\tilde{\tau}\tilde{\tau}}(x)}{f'(x)}dx&=\frac{c}{6L}\int_V\frac{\sin(\pi(R-x)/L)\sin(\pi(x+R)/L)}{\sin(2\pi R/L)}dx&\nonumber\\
&=-\frac{R c\cot(2\pi R/L)}{6L}+\frac{c}{12\pi}.
\end{align}

\subsection*{Finite temperature}\label{9.2}
 We calculate the entanglement entropy on a cylinder from the modular Hamiltonian. For the subsystem, $A=(-R, R)$  in an infinite line at finite temperature $1/\beta$, the stress tensor of the system is,
\begin{equation}
T_{\tilde{\tau}\tilde{\tau}}(x)=\frac{\pi c}{6\beta^2}.
\end{equation}
In the finite temperature case, the mapping function is chosen to be
\begin{equation}
f(z)=\log\frac{e^{2\pi z/ \beta}-e^{-2\pi R/\beta}}{e^{2\pi R/ \beta}-e^{-2\pi R/\beta}}.
\end{equation}
Then we obtain the modular Hamiltonian as
\begin{equation}
\mathcal{K}_A=\frac{\beta}{\pi}\int_A\frac{\sinh(\pi(R-x)/\beta)\sinh(\pi(R+x)/\beta)}{\sinh(2\pi R/\beta)}T_{00}(x)dx+\int_A\frac{c}{12\pi}\frac{\{f(z),z\}}{f'(z)}dz,
\end{equation}
and
\begin{equation}\label{temp}
\beta(x)=\frac{\beta}{\pi}\frac{\sinh(\pi(R-x)/\beta)\sinh(\pi(R+x)/\beta)}{\sinh(2\pi R/\beta)}.
\end{equation}
For the first term of the modular Hamiltonian, we obtain
\begin{equation}
\frac{\beta}{\pi}\int_A\frac{\sinh(\pi(R-x)/\beta)\sinh(\pi(R+x)/\beta)}{\sinh(2\pi R/\beta)}T_{00}(x)dx=\frac{\beta}{\pi}(R \coth\frac{2\pi R}{\beta}-\frac{\beta}{2\pi})(\frac{\pi c}{6\beta^2}).
\end{equation}
The second term of the modular Hamiltonian behaves as
\begin{equation}
\int_A\frac{c}{12\pi}\frac{\{f(z),z\}}{f'(z)}dz=\frac{c}{12\pi}(1-\frac{\pi R\cosh\frac{2\pi R}{\beta}}{\beta\sinh\frac{2\pi R}{\beta}}+\log(\frac{\beta}{\pi\epsilon}\sinh\frac{2\pi R}{\beta})).
\end{equation}
So the modular Hamiltonian on a cylinder at finite temperature is
\begin{equation}
\mathcal{K}_A=\frac{c}{12\pi}\log(\frac{\beta}{\pi\epsilon}\sinh(\frac{2\pi R}{\beta}))-\frac{cR}{12\beta}\coth\frac{2\pi R}{\beta}.
\end{equation}

With Wick's theorem, combine ~(\ref{i1}) with~(\ref{i2}) mentioned below to rewrite the equation~(\ref{Dlocal}) as
\begin{align}
\frac{{S_A^{local}}^{(1)}(\lambda)}{4\pi}&=-\frac{i\lambda}{2}\int_A\beta(x)\int_{\mathcal{M}}(\frac{\bar{\partial}\bar{P}_\nu(0)(\partial P_\nu(z-x))^2}{2\pi}-\frac{\bar\partial \bar{P}_\nu(0)\partial^2 P_\nu(z-x)P_\nu(z-x)}{2\pi})dxd^2z&\nonumber\\
&=-\lambda\int_A\beta(x)(\frac{\bar{\partial}\bar{P}_\nu(0)(e_{\nu-1}(\pi+i2\tau\eta_1)+(e^2_{\nu-1}-\frac{g_2}{6})(-i\tau))}{2\pi}&\nonumber\\
&\quad-\frac{\bar\partial \bar{P}_\nu(0)(-e_{\nu-1}(\pi+i2\tau\eta_1)-(e^2_{\nu-1}-\frac{g_2}{6})(-i\tau))}{2\pi})dx.
\end{align}
In the ``high temperature" limit $i\tau\to 0$, the correction of the entanglement entropy simplifies to
\begin{align}
\lim_{\beta\to 0}{S_A^{local}}^{(1)}(\lambda)&=-2\pi\lambda\lim_{\beta\to 0}\int_A\beta(x)(2\bar{\partial}\bar{P}_\nu(0)e_{\nu-1}
)dx&\nonumber\\
&=2\pi\lambda\lim_{\beta\to 0}\int_A\beta(x)\frac{16\pi^2\partial_\tau\partial_{\bar{\tau}}Z}{Z}dx.
\end{align}
It confirms our statement in the main text that only the second term contributes to the ``high temperature'' limit.

\section{Perturbative methods}\label{9.b}
To be self-consistent, we review briefly the perturbative methods for computing entanglement entropy (see (\ref{ET})) in terms of modular Hamiltonian as demonstrated in \cite{Rosenhaus:2014ula,moosa}.
Evaluating a theory under perturbation induced by operator $\mathcal{O}(x)$, the action of the perturbed theory is given by
\begin{equation}
\mathcal{I}_\lambda=\mathcal{I}_0+\lambda\mathcal{O},
\end{equation}
where
$
\mathcal{O}=\int d^dx\mathcal{O}(x),
$
and $\lambda$ is the coupling constant.

From eq.(\ref{es}), the entanglement entropy can be written in the path integral formalism as
\begin{equation}
    S_A(\lambda)=\frac{1}{Z_\lambda}\int D\psi e^{\mathcal{-I}_\lambda}\Big(2\pi\mathcal{K}_A(\lambda)+\log(\text{Tr}_Ae^{-2\pi\mathcal{K}_A})\Big),
\end{equation}
where the partion function is $Z_\lambda=\text{Tr}_Ae^{-2\pi\mathcal{K}_A(\lambda)}$.
It follows that
\be\ba\label{ssa}
\frac{dS_A(\lambda)}{d\lambda}=&\frac{1}{Z_\lambda}\int D\psi e^{\mathcal{-I}_\lambda}\Big(-\frac{d\mathcal{I}_\lambda}{d\lambda}(2\pi\mathcal{K}_A(\lambda)+\log(\text{Tr}_Ae^{-2\pi\mathcal{K}_A}))+\frac{d(2\pi\mathcal{K}_A(\lambda)+\log(\text{Tr}_Ae^{-2\pi\mathcal{K}_A}))}{d\lambda}\\
& -\frac{d\log Z_\lambda}{d\lambda}(2\pi\mathcal{K}_A(\lambda)+\log(\text{Tr}_Ae^{-2\pi\mathcal{K}_A}))\Big) \\
=&-\langle2\pi\mathcal{O}\mathcal{K}_A(\lambda)\rangle_\lambda+\langle\mathcal{O}\rangle_\lambda\langle2\pi\mathcal{K}_A(\lambda)\rangle_\lambda+\langle\frac{d(2\pi\mathcal{K}_A(\lambda)+\log(\text{Tr}_Ae^{-2\pi\mathcal{K}_A}))}{d\lambda}\rangle_\lambda,
\ea\ee
where we used
\be\ba
    \frac{d\mathcal{I}_\lambda}{d\lambda}=\mathcal{O}
,~~
    \frac{d\log Z_\lambda}{d\lambda}=-\langle\mathcal{O}\rangle_\lambda.
\ea\ee
Note that the last term in the last line of (\ref{ssa}) identically zero
\be\ba
 &\langle\frac{d(2\pi\mathcal{K}_A(\lambda)+\log(\text{Tr}_Ae^{-2\pi\mathcal{K}_A}))}{d\lambda}\rangle_\lambda\\
=&\text{Tr}_A\Big(\frac{e^{-2\pi\mathcal{K}_A}}{\text{Tr}_Ae^{-2\pi\mathcal{K}_A}}\frac{d(2\pi\mathcal{K}_A(\lambda)+\log(\text{Tr}_Ae^{-2\pi\mathcal{K}_A}))}{d\lambda}\Big)\\
=&-\frac{d}{d\lambda}\text{Tr}_A\frac{e^{-2\pi\mathcal{K}_A}}{\text{Tr}_Ae^{-2\pi\mathcal{K}_A}}=0
\ea\ee
Therefore we have
\begin{equation}\label{jcs1}
     \frac{dS_A(\lambda)}{d\lambda}=-\langle2\pi\mathcal{O}\mathcal{K}_A(\lambda)\rangle_\lambda+\langle\mathcal{O}\rangle_\lambda\langle2\pi\mathcal{K}_A(\lambda)\rangle_\lambda.
\end{equation}
Similarly, the second-order derivative of the entanglement entropy could also be worked out as
\begin{equation}
\frac{d^2S_A(\lambda)}{d^2\lambda}=\langle2\pi\mathcal{O}\mathcal{O}\mathcal{K}_A(\lambda)\rangle_\lambda-\langle\mathcal{O}\frac{d2\pi\mathcal{K}_A(\lambda)}{d\lambda}\rangle_\lambda-\langle\mathcal{O}\mathcal{O}\rangle_\lambda\langle2\pi\mathcal{K}_A(\lambda)\rangle_\lambda-\langle\mathcal{O}\rangle_\lambda\langle2\pi\mathcal{O}\mathcal{K}_A(\lambda)\rangle_\lambda.
\end{equation}

\section{Some useful integrals}\label{9.21}
In this section we introduce the Weierstrass functions and the related integrals which are used in the context.

From our convention, the torus $(T^2)$ is defined by the identificaiton on the complex plane $z\sim z+2\omega_1$ and $z\sim z+2\omega_2$ with $2\omega_1=1$, $2\omega_1=i\tau$.

The Weierstrass P function is defined as \cite{NIK}
\begin{equation}
 P(z)=\frac{1}{z^2}+\sum_{\{m,n\}\neq\{0,0\}}\Big(\frac{1}{(z-\tilde{\omega})^2}-\frac{1}{\tilde{\omega}^2}\Big), \quad \tilde{\omega}=2m\omega_1+2n\omega_2.
\end{equation}
The Weierstrass P function is elliptic function which is doubly periodic on the complex plane with periods $2\omega_1$ and $2\omega_2$.
The Laurent series expansion of $P(z)$ in the neighborhood of $z=0$ is
\begin{equation}
P(z)=\frac{1}{z^2}+\frac{g_2}{20}z^2+\frac{g_3}{28}z^4+O(z^6)
\end{equation}
where
\begin{equation}
    g_2=\sum_{\{m,n\}\neq\{0,0\}}\frac{60}{\tilde{\omega}^4},\quad g_3=\sum_{\{m,n\}\neq\{0,0\}}\frac{140}{\tilde{\omega}^6},
\end{equation}
and
\begin{equation}
e_1=P(\omega_1),\quad e_2=P(-\omega_1-\omega_2),\quad\lim_{\tau\to\infty}e_1=\frac{2\pi^2}{3},\quad \lim_{\tau\to\infty}e_2=\frac{-\pi^2}{3}.
\end{equation}

The primitive function of the Weierstrass $-P(z)$ function is the Weierstrass zeta-function $\zeta(z)$ which satisfies
\begin{equation}
-P(z)=\partial\zeta(z),
\end{equation}
where the  Weierstrass zeta-function $\zeta(z)$ is defined by
\begin{equation}
\zeta(z)=\frac{1}{z}+\sum_{\{m,n\}\neq\{0,0\}}\Big(\frac{1}{z-\tilde{\omega}}+\frac{1}{\tilde{\omega}}+\frac{z}{\omega^3}\Big)
\end{equation}

\subsection*{ The integrals for local modular Hamiltonian}\label{apc}
The calculation in subsection\ref{albb} involves integrals with singularity, we use the following strategy to regularize the result as in \cite{He:2020cxp}.
For double periodic meromorphic function $W(z)$, its integral is
\begin{align}\label{wer}
\int_{\mathcal{M}}d^2xW(z)&=\frac{i}{2}\int_{\mathcal{M}}W(z)dz\wedge d\bar{z}=\frac{i}{2}\int_{\mathcal{M}}d[(z-\bar{z})W(z)dz]=\frac{i}{2}\int_{\partial\mathcal{M}}(z-\bar{z})W(z) dz\nonumber\\
&=\frac{i}{2}\int^{z_0+2w_1}_{z_0}(z-\bar{z})W(z) dz+\frac{i}{2}\int^{z_0+2w_1+2w_2}_{z_0+2w_1}(z-\bar{z})W(z) dz+\frac{i}{2}\int^{z_0+2w_2}_{z_0+2w_1+2w_2}(z-\bar{z})W(z) dz&\nonumber\\
&\quad+\frac{i}{2}\int^{z_0}_{z_0+2w_2}(z-\bar{z})W(z) dz-\frac{i}{2}\lim_{r\to 0}\oint_{Poles}(z-\bar{z})W(z)dz&\nonumber\\
&=-\frac{i}{2}\int^{z_0+2w_1}_{z_0}(2w_2-2\bar{w}_2)W(z) dz+\frac{i}{2}\int^{z_0+2w_2}_{z_0}(z-\bar{z}+2w_1-2w_1)W(z+2w_1) dz&\nonumber\\
&\quad-\frac{i}{2}\int^{z_0+2w_2}_{z_0}(z-\bar{z})W(z) dz-\frac{i}{2}\lim_{r\to 0}\oint_{Poles}(z-\bar{z})W(z)dz&\nonumber\\
&=\tau_2\int^{z_0+2w_1}_{z_0}W(z)dz-\frac{i}{2}\lim_{r\to 0}\oint_{Poles}(z-\bar{z})W(z)dz.&
\end{align}
As an example, calculate the integral of Weierstrass P function which is
\begin{align}\label{c8}
\int_{T^2} d^2z P(z)&=\frac{-i}{2}\oint_{\partial T^2}\zeta(z)dz\nonumber\\
    &=-\frac{i}{2}\int^{z_0+\omega_1}_{z_0}d\bar{z}(\zeta(z)-\zeta(z+\omega_2))+\frac{i}{2}\int^{z_0+\omega_2}_{z_0}d\bar{z}(\zeta(z)-\zeta(z+\omega_1))\nonumber\\
    &=\pi-2\tau\eta_1.
\end{align}
In a similar manner, we can work out the following integrals which are needed in main text.

\begin{equation}\label{c9}
\int_{\mathcal{M}} d^2x \partial^2P(z)=0.
\end{equation}
\begin{equation}\label{i1}
\int_{\mathcal{M}}(\partial P_v(z))^2 d^2x=e_{\nu-1}(\pi+i2\tau\eta_1)+(e^2_{\nu-1}-\frac{g_2}{6})(-i\tau),
\end{equation}
\begin{equation}\label{i2}
\int_{\mathcal{M}}(P_\nu(z)\partial^2P_\nu(z))d^2x=-e_{\nu-1}(\pi+i2\tau\eta_1)-(e^2_{\nu-1}-\frac{g_2}{6})(-i\tau).
\end{equation}
where we used the fact that  elliptic fucntions  which can be written in terms of Weierstrass $P$ function and its derivatives, for example
\begin{equation}
(\partial P_\nu(z))^2=\frac{1}{6}\partial^2P(z)+e_{\nu-1}P(z)+e^2_{\nu-1}-\frac{g_2}{6}.
\end{equation}
For more detailed computation, please refer to \cite{He:2020cxp}.

\subsection*{The integrals for bi-local modular Hamiltonian}\label{9.4}
This subsection we introduce the integral which are used to calculate the integral for bi-local modular Hamiltonian in subsection\ref{eebd}.

The Jacobi$-\vartheta$ functions define as
\begin{align}
&\vartheta_1(z|\tau)=2q^{1/4}\sum_{n=0}^\infty q^{n(n+1)}\sin((2n+1)\pi z),\nonumber\\
&\vartheta_2(z|\tau)=2q^{1/4}\sum_{n=0}^\infty(-1)^nq^{n(n+1)}\cos((2n+1)\pi z).
\end{align}

Because $P_\nu$ and $\partial P_\nu$ are double periodic function, we use the regularization introduced in App. \ref{9.4}. Then we have
\begin{equation}\label{ptheta}
\int_{\mathcal{M}}\partial P_2(x_1-z)P_2(y-z)d^2x=-i\tau\int_0^1\partial P_2(x_1-z)P_2(y-z)dz-\frac{i}{2}\lim_{r\to0}\oint_{pole}(z-\bar{z})\partial P_2(x_1-z)P_2(y-z)dz.
\end{equation}
At the low temperature region $\beta\to\infty$, we obtain (\ref{wf})
\begin{equation}\label{lt}
\lim_{\beta\to\infty}P_2(z)=\frac{\pi\cos(\pi z)}{\sin(\pi z)}.
\end{equation}
Taking the equation (\ref{lt}) into the first term of  ~(\ref{ptheta}), we have the following results
\begin{equation}
\lim_{\beta\to \infty}\int_0^1\partial P_2(x_1-z)P_2(y-z)dz=\int^1_0\frac{\pi^3\cos(\pi(y-z))}{\sin^2(\pi(x_1-z))\sin(\pi(y-z))}dz=0.
\end{equation}
and
\begin{align}
&\quad-\frac{i}{2}\lim_{\beta\to  \infty}\lim_{r\to0}\oint_{pole}(z-\bar{z})\partial P_2(x_1-z)P_2(y-z)dz&\nonumber\\
&=-\frac{i}{2}\lim_{\beta\to  \infty}\lim_{|z-y|=r}\oint_{pole}(z-\bar{z})\partial P_2(x_1-z)P_2(y-z)dz-\frac{i}{2}\lim_{\beta\to  \infty}\lim_{|z-x|=r}\oint_{pole}(z-\bar{z})\partial P_2(x_1-z)P_2(y-z)dz&\nonumber\\
&=\frac{\pi^3\cos(\pi(x_1-y))}{\sin(\pi(x_1-y))}=\pi^3\lim_{\beta\to\infty}P_2(x_1-y).&
\end{align}

Thus we obtain
\begin{equation}\label{pearl}
\lim_{\beta\to\infty}\int_{\mathcal{M}}\partial P_2(x_1-z)P_2(y-z)d^2x=\pi^3\lim_{\beta\to\infty}P_2(x_1-y).
\end{equation}

\subsection*{The integrals in the half-line}\label{c.3}
In this part we calculate the eq.(\ref{wuyu}) in Sec.\ref{e}.
\begin{align}\label{ere}
\langle\mathcal{T}^{(0)}_T(\omega)\frac{d\mathcal{K}_P(\lambda)}{d\lambda}\rangle_0&=\frac{i}{2}\int_\mathcal{L}\int_P\pi c^2x\Big(\frac{1}{3(\omega-v)^5(\bar{\omega}-\bar{v})^3}+\frac{1}{3(\omega-v)^3(\bar{\omega}-\bar{v})^5}\Big)dxd^2\omega\nonumber\\
&=\frac{i}{2}\int_\mathcal{L}\int_P\frac{\pi c^2x}{3}\Big(\frac{1}{(\omega-x)^3(\bar{\omega}-x)^3}\Big)\Big(\frac{1}{(\omega-x)^2}+\frac{1}{(\bar{\omega}-x)^2}\Big)dxd^2\omega\nonumber\\
&=\frac{i}{2}\int_\mathcal{L}\int_P\frac{\pi c^2x}{3}\Big(\frac{2}{(y-x+i\tilde{\tau})^3(y-x-i\tilde{\tau})^3}\Big)\frac{(y-x)^2-\tilde{\tau}^2}{(\tilde{\tau}^2+(y-x)^2)^2}dxd^2\omega\nonumber\\
&=-\int_\mathcal{L}\int_P\frac{2\pi c^2x}{3}\frac{\cos(2\theta)}{r^{7}}dxdrd\theta\nonumber\\
&=0.
\end{align}
where the coordinate transformation $\frac{i}{2}d^2\omega=dyd\tilde{\tau}=rdrd\theta$ is used and we drop out the contact terms during the calculation.

To calculate eq.(\ref{ga}), we have that
\begin{equation}
\langle\mathcal{T}^{(0)}_T\mathcal{T}^{(0)}_T\mathcal{K}_P\rangle_0=-\frac{1}{4}\int_{\mathcal{L}_1}\int_{\mathcal{L}_2}\int_Pd^2z_1d^2z_2dx\langle xT^{(0)}(z_1)\bar{T}^{(0)}(\bar{z_1})T^{(0)}(z_2)\bar{T}^{(0)}(\bar{z_2})(T^{(0)}(z)+\bar{T}^{(0)}(\bar{z}))\rangle_0
\end{equation}
For the first term of RHS integral, we have
\begin{align}
&\quad-\frac{1}{4}\int_{\mathcal{L}_1}\int_{\mathcal{L}_2}\int_Pd^2z_1d^2z_2dx\langle xT^{(0)}(z_1)\bar{T}^{(0)}(\bar{z_1})T^{(0)}(z_2)\bar{T}^{(0)}(\bar{z_2})(T^{(0)}(z))\rangle_0\nonumber\\
&=-\frac{1}{16}\int_{\mathcal{L}_1}\int_{\mathcal{L}_2}\int_Pd^2z_1d^2z_2dx\frac{xc^2}{(z_1-z_2)^2(z_1-z)^2(z_2-z)^2(\bar{z}_1-\bar{z}_2)^4}\nonumber\\
&=\frac{1}{48}\int_{\mathcal{L}_1}\int_{\mathcal{L}_2}\int_Pd^2z_1d^2z_2dx\frac{xc^2}{(z_1-z_2)^2(z_1-z)^2(z_2-z)^2}\partial_{\bar{z}_1}\frac{1}{(\bar{z}_1-\bar{z}_2)^3}\nonumber\\
&=-\frac{1}{48}\int_{\mathcal{L}_1}\int_{\mathcal{L}_2}\int_Pd^2z_1d^2z_2dx\partial_{\bar{z}_1}\Big(\frac{xc^2}{(z_1-z_2)^2(z_1-z)^2(z_2-z)^2}\Big)\frac{1}{(\bar{z}_1-\bar{z}_2)^3}\nonumber\\
&=\frac{\pi}{48}\int_{\mathcal{L}_1}\int_{\mathcal{L}_2}\int_Pd^2z_1d^2z_2dx\partial_{z_1}\delta^{(2)}(z_1-z_2)\frac{xc^2}{(z_1-z)^2(z_2-z)^2}\frac{1}{(\bar{z}_1-\bar{z}_2)^3}+(z_2\leftrightarrow z)\nonumber\\
&=\frac{\pi}{24}\int_{\mathcal{L}_1}\int_{\mathcal{L}_2}\int_Pd^2z_1d^2z_2dx\delta^{(2)}(z_1-z_2)\frac{xc^2}{(z_1-z)^3(z_2-z)^2}\frac{1}{(\bar{z}_1-\bar{z}_2)^3}+(z_2\leftrightarrow z)\nonumber\\
&=\frac{i\pi}{12}\int_{\mathcal{L}_2}\int_Pd^2z_2dx\frac{xc^2}{(z_2-z)^5}\frac{1}{(\bar{z}-\bar{z}_2)^3},
\end{align}
where we use $\partial{z}\frac{1}{\bar{z}}=\pi\delta^{(2)}(z)$ and get rid of the divergent terms when two points coincide.

Similarly, we also obtain
\begin{align}
&\quad-\frac{1}{4}\int_{\mathcal{L}_1}\int_{\mathcal{L}_2}\int_Pd^2z_1d^2z_2dx\Big(xT^{(0)}(z_1)\bar{T}^{(0)}(\bar{z_1})T^{(0)}(z_2)\bar{T}^{(0)}(\bar{z_2})(T^{(0)}(z))\Big)\nonumber\\
&=\frac{i\pi}{12}\int_{\mathcal{L}_2}\int_Pd^2z_2dx\frac{xc^2}{(\bar{z}_2-\bar{z})^5}\frac{1}{(z-z_2)^3}.
\end{align}
Thus we have
\begin{align}
\langle\mathcal{T}^{(0)}_T\mathcal{T}^{(0)}_T\mathcal{K}_P\rangle_0&=\frac{ic^2\pi}{12}\int_{\mathcal{L}_2}\int_Pd^2z_2dx\Big(\frac{x}{(\bar{z}_2-\bar{z})^5}\frac{1}{(z_2-z)^3}+\frac{x}{(z_2-z)^5}\frac{1}{(\bar{z}_2-\bar{z})^3}\Big)\nonumber\\
&=0,
\end{align}
where the eq.(\ref{ere}) is used in the last step.

\subsection*{The integrals of mass deformed fermions}\label{9.5}
In the appendix we evaluate the term $\langle\mathcal{T}_m\mathcal{T}_m\rangle$ appearing in section~(\ref{mee}).
From~(\ref{weyl}), we have
\begin{equation}\label{ttm2}
\langle\mathcal{T}_m\mathcal{T}_m\rangle_0\sim\int_{\mathcal{M}_1}\int_{\mathcal{M}_2}dx_1^2dx_2^2P_\nu(z_1-z_2)\bar{P}_\nu(\bar{z}_1-\bar{z}_2).
\end{equation}
To calculate it, we first analyze the first integral $\int_{\mathcal{M}_1}d^2x_1P_\nu(z_1-z_2)\bar{P}_\nu(\bar{z}_1-\bar{z}_2)$. 
According to the definition of $P_\nu$ function in~(\ref{wf}), we have
\begin{equation}
P_\nu(z)\bar{P}_\nu(\bar{z})=\lvert P(z)-e_{\nu-1}\rvert.
\end{equation}
As the Weierstrass $P(z)$ is double periodic function, so is the function $P_\nu(z)\bar{P}_\nu(\bar{z})$.

Thus we obtain
\begin{equation}\label{wired}
    \int_{\mathcal{M}_1}d^2x_1P_\nu(z_1-z_2)\bar{P}_\nu(\bar{z}_1-\bar{z}_2)=\int_{\mathcal{M}_1}d^2x_1P_\nu(z_1)\bar{P}_\nu(\bar{z}_1)=C.
\end{equation}
where the factor $C$ is a undetermined constant which implies that the integral is independent of the parameter $z_2$. To have a glimpse of the integral, we analyze it in the $\nu=2$ sector the low temperature limit $\beta\to\infty$.
Taking~(\ref{lt}) into~(\ref{codozz}), the result is
\small{
\begin{align}\label{stro}
&\quad\lim_{\beta\to \infty}\int_{\mathcal{M}_1}\int_{\mathcal{M}_2}dx_1^2dx_2^2P_2(z_1-z_2)\bar{P}_2(\bar{z}_1-\bar{z}_2)\nonumber&\\
&=\int_{\mathcal{M}_1}\int_{\mathcal{M}_2}\frac{\pi^2 \cos(\pi(z_1-z_2))}{\sin(\pi(z_1-z_2))}\frac{ \cos(\pi(\bar{z}_1-\bar{z}_2))}{\sin(\pi(\bar{z}_1-\bar{z}_2))}dx_1^2dx_2^2\nonumber&\\
&=-\int_{\mathcal{M}_1}\int_{\mathcal{M}_2}\partial_{z_2}\Big(\ln(\sin(\pi(z_1-z_2))\frac{\pi\cos(\pi(\bar{z}_1-\bar{z}_2))}{\sin(\pi(\bar{z}_1-\bar{z}_2))}\Big)+\int_{\mathcal{M}_1}\int_{\mathcal{M}_2}\ln(\sin(\pi(z_1-z_2))\partial_{z_2}(\frac{ \cos(\pi(\bar{z}_1-\bar{z}_2))}{\sin(\pi(\bar{z}_1-\bar{z}_2))})\nonumber&\\
&=-\frac{i}{2}\int_{\mathcal{M}_1}\int_{\partial \mathcal{M}_2}d\bar{z}_2\ln(\sin(\pi(z_1-z_2))\frac{\pi\cos(\pi(\bar{z}_1-\bar{z}_2))}{\sin(\pi(\bar{z}_1-\bar{z}_2))}+\frac{i}{2}\int_{\mathcal{M}_1}\int_{\partial {pole}_2}d\bar{z}_2\ln(\sin(\pi(z_1-z_2))\frac{\pi\cos(\pi(\bar{z}_1-\bar{z}_2))}{\sin(\pi(\bar{z}_1-\bar{z}_2))}\nonumber&\\
&\quad+\int_{\mathcal{M}_1}\int_{\mathcal{M}_2}\ln(\sin(\pi(z_1-z_2))\delta(z_1-z_2)\cos(\pi(\bar{z}_1-\bar{z}_2))\nonumber&\\
&=-\frac{i}{2}\int_{\mathcal{M}_1}\int_{\partial\mathcal{M}_2}d\bar{z}_2\ln(\sin(\pi(z_1-z_2))\frac{\pi\cos(\pi(\bar{z}_1-\bar{z}_2))}{\sin(\pi(\bar{z}_1-\bar{z}_2))}+\lim_{r\to 0}\beta\pi\ln\pi r+i\beta\pi^2\nonumber&\\
&=-\frac{i}{2}\int_{\mathcal{M}_1}(\int^1_0+\int^{1+\beta}_1+\int^{\beta}_{1+\beta}+\int^0_\beta)\ln(\sin(\pi(z_1-z_2))\frac{\pi\cos(\pi(\bar{z}_1-\bar{z}_2))}{\sin(\pi(\bar{z}_1-\bar{z}_2))}d\bar{z}_2+\lim_{r\to 0}\beta\pi\ln\pi r+i\beta\pi^2\nonumber&\\
&=\beta\pi\ln\pi,&
\end{align}
}
in the above integral we have used $\partial{z}\frac{1}{\bar{z}}=\pi\delta^{(2)}(z)$ and omit both the divergent term and imaginary term in the calculation. In the last step of the equation, $d\bar{z}=dre^{-i\theta}=e^{-i\theta}dz$ is used when the angular coordinate $\theta=0,\frac{\pi}{2}$ is fixed for the integral path we choose. Equation (\ref{wired}) implicating that the integral in the domain $\mathcal{M}_2$ is independent of the location of $z_1$, we may take $z_1=0$ in the calculation.

Here we calculate the integral in eq.(\ref{stran}) for the system at the low temperature limit $\beta\to\infty$ in $\nu=2$ sector.
\begin{align}
 &\lim_{\beta\to \infty}\int_{\mathcal{M}_1}\int_{\mathcal{M}_2}\Big(\frac{\partial P_2(z-z_2)P_2(z-z_1)\bar{P}_2(\bar{z}_1-\bar{z}_2)}{2\pi^2}\Big)dx_1^2d{x_2}^2\nonumber\\
  =&\lim_{\beta\to \infty}\int_{\mathcal{M'}_1}\int_{\mathcal{M}_2}\Big(\frac{\partial P_2(z-z_2)P_2(z-z'_1-z_2)\bar{P}_2(\bar{z}'_1)}{2\pi^2}\Big)d{x'_1}^2d{x_2}^2\nonumber\\
   =&\lim_{\beta\to \infty}\int_{\mathcal{M'}_1}\Big(\frac{\partial P_2(z-z_2)P_2(z-z'_1-z_2)\bar{P}_2(\bar{z}'_1)}{2}\Big)d{x'_1}^2d{x_2}^2\nonumber\\
   =&\lim_{\beta\to \infty}\int_{\mathcal{M'}_1}\Big(\frac{\pi^3P_2(z'_1)\bar{P}_2(\bar{z}'_1)}{2}\Big)d{x'_1}^2\nonumber\\
   =&\frac{\pi^4\ln\pi}{2},
\end{align}
   where we seperately use the eq.(\ref{pearl}) and eq.(\ref{stro}) in the third line and last line.


\begin{thebibliography}{99}

\bibitem{Calabrese:2004eu}
P.~Calabrese and J.~L.~Cardy,
``Entanglement entropy and quantum field theory,''
J. Stat. Mech. \textbf{0406}, P06002 (2004)
\href{https://arxiv.org/pdf/hep-th/0405152.pdf}{[arXiv:hep-th/0405152 [hep-th]]}.

\bibitem{Ryu:2006bv}
S.~Ryu and T.~Takayanagi,
``Holographic derivation of entanglement entropy from AdS/CFT,''
Phys. Rev. Lett. \textbf{96}, 181602 (2006)
\href{https://arxiv.org/pdf/hep-th/0605073.pdf}{[arXiv:hep-th/0603001 [hep-th]]}.

\bibitem{Hubeny:2007xt}
V.~E.~Hubeny, M.~Rangamani and T.~Takayanagi,
``A Covariant holographic entanglement entropy proposal,''
JHEP \textbf{07}, 062 (2007)
\href{https://arxiv.org/pdf/0705.0016.pdf}{[arXiv:0705.0016 [hep-th]]}.

\bibitem{Lashkari:2015dia}
N.~Lashkari,
``Modular Hamiltonian for Excited States in Conformal Field Theory,''
Phys. Rev. Lett. \textbf{117}, no.4, 041601 (2016)
\href{https://arxiv.org/pdf/1508.03506.pdf}{[arXiv:1508.03506 [hep-th]]}.

\bibitem{Sarosi:2017rsq}
G.~S\'arosi and T.~Ugajin,
``Modular Hamiltonians of excited states, OPE blocks and emergent bulk fields,''
JHEP \textbf{01}, 012 (2018)
\href{https://arxiv.org/pdf/1705.01486.pdf}{[arXiv:1705.01486 [hep-th]]}.

\bibitem{Alcaraz:2011tn}
F.~C.~Alcaraz, M.~I.~Berganza and G.~Sierra,
``Entanglement of low-energy excitations in Conformal Field Theory,''
Phys. Rev. Lett. \textbf{106}, 201601 (2011)
\href{https://arxiv.org/pdf/1101.2881.pdf}{[arXiv:1101.2881 [cond-mat.stat-mech]]}.

\bibitem{Bhattacharya:2012mi}
J.~Bhattacharya, M.~Nozaki, T.~Takayanagi and T.~Ugajin,
``Thermodynamical Property of Entanglement Entropy for Excited States,''
Phys. Rev. Lett. \textbf{110}, no.9, 091602 (2013)
\href{https://arxiv.org/pdf/1212.1164.pdf}{[arXiv:1212.1164 [hep-th]]}.



\bibitem{Nozaki:2013vta}
M.~Nozaki, T.~Numasawa, A.~Prudenziati and T.~Takayanagi,
``Dynamics of Entanglement Entropy from Einstein Equation,''
Phys. Rev. D \textbf{88} (2013) no.2, 026012
\href{https://arxiv.org/pdf/1304.7100.pdf}{[arXiv:1304.7100 [hep-th]]}.

\bibitem{Guo:2013aca}
W.~z.~Guo, S.~He and J.~Tao,
``Note on Entanglement Temperature for Low Thermal Excited States in Higher Derivative Gravity,''
JHEP \textbf{08}, 050 (2013)
\href{https://arxiv.org/pdf/1305.2682.pdf}{[arXiv:1305.2682 [hep-th]]}.

\bibitem{Allahbakhshi:2013rda}
D.~Allahbakhshi, M.~Alishahiha and A.~Naseh,
``Entanglement Thermodynamics,''
JHEP \textbf{08} (2013), 102
\href{https://arxiv.org/pdf/1305.2728.pdf}{[arXiv:1305.2728 [hep-th]]}.

\bibitem{Blanco:2013joa}
D.~D.~Blanco, H.~Casini, L.~Y.~Hung and R.~C.~Myers,
``Relative Entropy and Holography,''
JHEP \textbf{08}, 060 (2013)
\href{https://arxiv.org/pdf/1305.3182.pdf}{[arXiv:1305.3182 [hep-th]]}.

\bibitem{Faulkner:2013ica}
T.~Faulkner, M.~Guica, T.~Hartman, R.~C.~Myers and M.~Van Raamsdonk,
``Gravitation from Entanglement in Holographic CFTs,''
JHEP \textbf{03}, 051 (2014)
\href{https://arxiv.org/pdf/1312.7856.pdf}{[arXiv:1312.7856 [hep-th]]}.

\bibitem{Faulkner:2016mzt}
T.~Faulkner, R.~G.~Leigh, O.~Parrikar and H.~Wang,
``Modular Hamiltonians for Deformed Half-Spaces and the Averaged Null Energy Condition,''
JHEP \textbf{09}, 038 (2016)
\href{https://arxiv.org/pdf/1605.08072.pdf}{[arXiv:1605.08072 [hep-th]]}.

\bibitem{Jafferis:2014lza}
D.~L.~Jafferis and S.~J.~Suh,
``The Gravity Duals of Modular Hamiltonians,''
JHEP \textbf{09}, 068 (2016)
\href{https://arxiv.org/pdf/1412.8465.pdf}{[arXiv:1412.8465 [hep-th]]}.

\bibitem{Jafferis:2015del}
D.~L.~Jafferis, A.~Lewkowycz, J.~Maldacena and S.~J.~Suh,
``Relative entropy equals bulk relative entropy,''
JHEP \textbf{06}, 004 (2016)
\href{https://arxiv.org/pdf/1512.06431.pdf}{[arXiv:1512.06431 [hep-th]]}.

\bibitem{Faulkner:2018faa}
T.~Faulkner, M.~Li and H.~Wang,
``A modular toolkit for bulk reconstruction,''
JHEP \textbf{04}, 119 (2019)
\href{https://arxiv.org/pdf/1806.10560.pdf}{[arXiv:1806.10560 [hep-th]]}.

\bibitem{Casini:2011kv}
H.~Casini, M.~Huerta and R.~C.~Myers,
``Towards a derivation of holographic entanglement entropy,''
JHEP \textbf{05}, 036 (2011)
\href{https://arxiv.org/pdf/1102.0440.pdf}{[arXiv:1102.0440 [hep-th]]}.

\bibitem{Bisognano:1975ih}
J.~J.~Bisognano and E.~H.~Wichmann,
``On the Duality Condition for a Hermitian Scalar Field,''
\href{https://aip.scitation.org/doi/10.1063/1.522605}{J. Math. Phys. \textbf{16}, 985-1007 (1975)}.

\bibitem{Bisognano:1976za}
J.~J.~Bisognano and E.~H.~Wichmann,
``On the Duality Condition for Quantum Fields,''
\href{https://aip.scitation.org/doi/abs/10.1063/1.522898}{J. Math. Phys. \textbf{17}, 303-321 (1976)}.

\bibitem{Wong:2013gua}
G.~Wong, I.~Klich, L.~A.~Pando Zayas and D.~Vaman,
``Entanglement Temperature and Entanglement Entropy of Excited States,''
JHEP \textbf{12}, 020 (2013)
\href{https://arxiv.org/pdf/1305.3291.pdf}{[arXiv:1305.3291 [hep-th]]}.

\bibitem{Cardy:2016fqc}
J.~Cardy and E.~Tonni,
``Entanglement Hamiltonians in two-dimensional conformal field theory,''
J. Stat. Mech. \textbf{1612}, no.12, 123103 (2016)
 \href{https://arxiv.org/pdf/1608.01283.pdf}{[arXiv:1608.01283 [cond-mat.stat-mech]]}.

\bibitem{Casini:2009vk}
H.~Casini and M.~Huerta,
``Reduced density matrix and internal dynamics for multicomponent regions,''
Class. Quant. Grav. \textbf{26}, 185005 (2009)
\href{https://arxiv.org/pdf/0903.5284.pdf}{[arXiv:0903.5284 [hep-th]]}.

\bibitem{Mintchev:2020uom}
M.~Mintchev and E.~Tonni,
``Modular Hamiltonians for the massless Dirac field in the presence of a boundary,''
JHEP \textbf{03}, 204 (2021)
\href{https://arxiv.org/pdf/2012.00703 .pdf}{[arXiv:2012.00703  [hep-th]]}.

\bibitem{Fries:2019ozf}
P.~Fries and I.~A.~Reyes,
``Entanglement Spectrum of Chiral Fermions on the Torus,''
Phys. Rev. Lett. \textbf{123}, no.21, 211603 (2019)
\href{https://arxiv.org/pdf/1905.05768.pdf}{[arXiv:1905.05768 [hep-th]]}.

\bibitem{Rosenhaus:2014woa}
V.~Rosenhaus and M.~Smolkin,
``Entanglement Entropy: A Perturbative Calculation,''
JHEP \textbf{12}, 179 (2014)
 \href{https://arxiv.org/pdf/1403.3733.pdf}{[arXiv:1403.3733 [hep-th]]}.

\bibitem{Balakrishnan:2017bjg}
S.~Balakrishnan, T.~Faulkner, Z.~U.~Khandker and H.~Wang,
``A General Proof of the Quantum Null Energy Condition,''
JHEP \textbf{09}, 020 (2019)
 \href{https://arxiv.org/pdf/1706.09432.pdf}{[arXiv:1706.09432 [hep-th]]}.

\bibitem{Faulkner:2015csl}
T.~Faulkner, R.~G.~Leigh and O.~Parrikar,
``Shape Dependence of Entanglement Entropy in Conformal Field Theories,''
JHEP \textbf{04}, 088 (2016)
\href{https://arxiv.org/pdf/1511.05179.pdf}{[arXiv:1511.05179 [hep-th]]}.

\bibitem{Allais:2014ata}
A.~Allais and M.~Mezei,
``Some results on the shape dependence of entanglement and R\'enyi entropies,''
Phys. Rev. D \textbf{91}, no.4, 046002 (2015)
\href{https://arxiv.org/pdf/1407.7249.pdf}{[arXiv:1407.7249 [hep-th]]}.

\bibitem{Rosenhaus:2014zza}
V.~Rosenhaus and M.~Smolkin,
``Entanglement Entropy for Relevant and Geometric Perturbations,''
JHEP \textbf{02}, 015 (2015)
\href{https://arxiv.org/pdf/1410.6530.pdf}{[arXiv:1410.6530 [hep-th]]}.

\bibitem{Balakrishnan:2020lbp}
S.~Balakrishnan and O.~Parrikar,
``Modular Hamiltonians for Euclidean Path Integral States,''
\href{https://arxiv.org/pdf/2002.00018.pdf}{[arXiv:2002.00018  [hep-th]]}.

\bibitem{Arias:2020qpg}
R.~Arias, M.~Botta-Cantcheff, P.~J.~Martinez and J.~F.~Zarate,
``Modular Hamiltonian for holographic excited states,''
Phys. Rev. D \textbf{102}, no.2, 026021 (2020)
 \href{https://arxiv.org/pdf/2002.04637.pdf}{[arXiv:2002.04637  [hep-th]]}.

\bibitem{Lashkari:2018oke}
N.~Lashkari, H.~Liu and S.~Rajagopal,
``Modular flow of excited states,''
JHEP \textbf{09}, 166 (2021)
\href{https://arxiv.org/pdf/1811.05052.pdf}{[arXiv:1811.05052  [hep-th]]}.

\bibitem{Cavaglia:2016oda}
A.~Cavagli\`a, S.~Negro, I.~M.~Sz\'ecs\'enyi and R.~Tateo,
``$T \bar{T}$-deformed 2D Quantum Field Theories,''
JHEP \textbf{10}, 112 (2016)
\href{https://arxiv.org/pdf/1608.05534.pdf}{[arXiv:1608.05534 [hep-th]]}.

\bibitem{Smirnov:2016lqw}
F.~A.~Smirnov and A.~B.~Zamolodchikov,
``On space of integrable quantum field theories,''
Nucl. Phys. B \textbf{915}, 363-383 (2017)
\href{https://arxiv.org/pdf/1608.05499.pdf}{[arXiv:1608.05499 [hep-th]]}.

\bibitem{McGough:2016lol}
L.~McGough, M.~Mezei and H.~Verlinde,
``Moving the CFT into the bulk with $ T\overline{T} $,''
JHEP \textbf{04}, 010 (2018)
\href{https://arxiv.org/pdf/1611.03470.pdf}{[arXiv:1611.03470 [hep-th]]}.

\bibitem{Kraus:2022mnu}
P.~Kraus, R.~Monten and K.~Roumpedakis,
``Refining the Cutoff 3d Gravity / $T\bar{T}$ Correspondence,''
\href{https://arxiv.org/pdf/2206.00674.pdf}{[arXiv:2206.00674 [hep-th]]}.

\bibitem{Donnelly:2018bef}
W.~Donnelly and V.~Shyam,
``Entanglement entropy and $T \overline{T}$ deformation,''
Phys. Rev. Lett. \textbf{121}, no.13, 131602 (2018)
\href{https://arxiv.org/pdf/1806.07444.pdf}{[arXiv:1806.07444 [hep-th]]}.

\bibitem{Chen:2018eqk}
B.~Chen, L.~Chen and P.~X.~Hao,
``Entanglement entropy in $T\overline{T}$-deformed CFT,''
Phys. Rev. D \textbf{98}, no.8, 086025 (2018)
\href{https://arxiv.org/pdf/1807.08293.pdf}{[arXiv:1807.08293 [hep-th]]}.

\bibitem{Ota:2019yfe}
T.~Ota,
``Comments on holographic entanglements in cutoff AdS,''
\href{https://arxiv.org/pdf/1904.06930.pdf}{[arXiv:1904.06930 [hep-th]]}.

\bibitem{Donnelly:2019pie}
W.~Donnelly, E.~LePage, Y.~Y.~Li, A.~Pereira and V.~Shyam,
``Quantum corrections to finite radius holography and holographic entanglement entropy,''
JHEP \textbf{05} (2020), 006
\href{https://arxiv.org/pdf/1909.11402.pdf}{[arXiv:1909.11402 [hep-th]]}.



\bibitem{Allameh:2021moy}
K.~Allameh, A.~F.~Astaneh and A.~Hassanzadeh,
``Aspects of holographic entanglement entropy for TT\textasciimacron{}-deformed CFTs,''
Phys. Lett. B \textbf{826} (2022), 136914
\href{https://arxiv.org/pdf/2111.11338.pdf}{[arXiv:2111.11338 [hep-th]]}.

\bibitem{Setare:2022qls}
M.~R.~Setare and S.~N.~Sajadi,
``Holographic Entanglement Entropy in $T\bar{T}$-deformed CFTs,''
\href{https://arxiv.org/pdf/2203.16445.pdf}{[arXiv:2203.16445 [hep-th]]}.

\bibitem{Jeong:2019ylz}
H.~S.~Jeong, K.~Y.~Kim and M.~Nishida,
``Entanglement and R\'enyi entropy of multiple intervals in $T\overline{T}$-deformed CFT and holography,''
Phys. Rev. D \textbf{100} (2019) no.10, 106015
\href{https://arxiv.org/pdf/1906.03894.pdf}{[arXiv:1906.03894 [hep-th]]}.

\bibitem{He:2019vzf}
S.~He and H.~Shu,
``Correlation functions, entanglement and chaos in the $ T\overline{T}/J\overline{T} $-deformed CFTs,''
JHEP \textbf{02} (2020), 088
\href{https://arxiv.org/pdf/1907.12603.pdf}{[arXiv:1907.12603 [hep-th]]}.

\bibitem{He:2020qcs}
S.~He,
``Note on higher-point correlation functions of the $T\bar T$ or $J\bar T$ deformed CFTs,''
Sci. China Phys. Mech. Astron. \textbf{64} (2021) no.9, 291011
\href{https://arxiv.org/pdf/2012.06202.pdf}{[arXiv:2012.06202 [hep-th]]}.

\bibitem{Asrat:2020uib}
M.~Asrat and J.~Kudler-Flam,
``$T\bar{T}$, the entanglement wedge cross section, and the breakdown of the split property,''
Phys. Rev. D \textbf{102} (2020) no.4, 045009
\href{https://arxiv.org/pdf/2005.08972.pdf}{[arXiv:2005.08972 [hep-th]]}.

\bibitem{Cardona:2022cmh}
B.~Cardona and J.~Molina-Vilaplana,
``Entanglement Renormalization of a $T\bar{T}$-deformed CFT,''
\href{https://arxiv.org/pdf/2203.00319.pdf}{[arXiv:2203.00319 [hep-th]]}.

\bibitem{Ageev:2022kpm}
D.~S.~Ageev, A.~I.~Belokon and V.~V.~Pushkarev,
``From locality to irregularity: Introducing local quenches in massive scalar field theory,''
\href{https://arxiv.org/pdf/2205.12290.pdf}{[arXiv:2205.12290 [hep-th]]}.

\bibitem{Guica:2020uhm}
M.~Guica and R.~Monten,
``Infinite pseudo-conformal symmetries of classical $T \bar T$, $J \bar T $ and $J T_a$ - deformed CFTs,''
SciPost Phys. \textbf{11}, 078 (2021)
\href{https://arxiv.org/pdf/2011.05445.pdf}{[arXiv:2011.05445 [hep-th]]}.

\bibitem{Kraus:2021cwf}
P.~Kraus, R.~Monten and R.~M.~Myers,
``3D Gravity in a Box,''
SciPost Phys. \textbf{11}, 070 (2021)
\href{https://arxiv.org/pdf/2103.13398.pdf}{[arXiv:2103.13398 [hep-th]]}.

\bibitem{He:2021bhj}
M.~He, S.~He and Y.~h.~Gao,
``Surface charges in Chern-Simons gravity with $ T\overline{T} $ deformation,''
JHEP \textbf{03}, 044 (2022)
\href{https://arxiv.org/pdf/2109.12885.pdf}{[arXiv:2109.12885 [hep-th]]}.

\bibitem{Rangamani:2015agy}
M.~Rangamani, M.~Rozali and A.~Vincart-Emard,
``Dynamics of Holographic Entanglement Entropy Following a Local Quench,''
JHEP \textbf{04}, 069 (2016)
\href{https://arxiv.org/pdf/1512.03478.pdf}{[arXiv:1512.03478 [hep-th]]}.

\bibitem{Fries:2019acy}
P.~Fries and I.~A.~Reyes,
``Entanglement and relative entropy of a chiral fermion on the torus,''
Phys. Rev. D \textbf{100}, no.10, 105015 (2019)
\href{https://arxiv.org/pdf/1906.02207.pdf}{[arXiv:1906.02207 [hep-th]]}.

\bibitem{moosa}
M.~Moosa
\href{http://www.sms.edu.pk/wp-content/uploads/2018/03/Entanglement-Entropy.pdf}{``Entanglement in QFT and Holography''.}

\bibitem{Rosenhaus:2014ula}
V.~Rosenhaus and M.~Smolkin,
``Entanglement entropy, planar surfaces, and spectral functions,''
JHEP \textbf{09} (2014), 119
\href{https://arxiv.org/pdf/1407.2891.pdf}{[arXiv:1407.2891 [hep-th]]}.

\bibitem{Herzog:2013py}
C.~P.~Herzog and T.~Nishioka,
``Entanglement Entropy of a Massive Fermion on a Torus,''
JHEP \textbf{03}, 077 (2013)
\href{https://arxiv.org/pdf/1301.0336.pdf}{[arXiv:1301.0336 [hep-th]]}.

\bibitem{Nozaki:2015mca}
M.~Nozaki, T.~Numasawa and S.~Matsuura,
``Quantum Entanglement of Fermionic Local Operators,''
JHEP \textbf{02}, 150 (2016)
\href{https://arxiv.org/pdf/1507.04352.pdf}{[arXiv:1507.04352 [hep-th]]}.

\bibitem{Akal:2021foz}
I.~Akal, Y.~Kusuki, N.~Shiba, T.~Takayanagi and Z.~Wei,
``Holographic moving mirrors,''
Class. Quant. Grav. \textbf{38}, no.22, 224001 (2021)
\href{https://arxiv.org/pdf/2106.11179.pdf}{[arXiv:2106.11179 [hep-th]]}.

\bibitem{WanMokhtar:2018lwi}
W.~M.~H.~Wan Mokhtar,
``Radiation from a receding mirror: Unruh-DeWitt detector distinguishes a Dirac fermion from a scalar boson,''
Class. Quant. Grav. \textbf{37}, no.7, 075011 (2020)
\href{https://arxiv.org/pdf/1806.11511.pdf}{[arXiv:1806.11511 [gr-qc]]}.

\bibitem{Inch}
I.~Peschel,
``Calculation of reduced density matrices from
correlation functions,''
Journal of Physics A: Mathematical and General
\href{https://arxiv.org/pdf/cond-mat/0212631.pdf}{[arXiv:0212631[cond-mat]]}.

\bibitem{Blanco:2019cet}
D.~Blanco, A.~Garbarz and G.~P\'erez-Nadal,
``Entanglement of a chiral fermion on the torus,''
JHEP \textbf{09}, 076 (2019)
\href{https://arxiv.org/pdf/1906.07057 .pdf}{[arXiv:1906.07057  [hep-th]]}.


\bibitem{DiFrancesco:1997nk}
P.~Di Francesco, P.~Mathieu and D.~Senechal,
``Conformal Field Theory,''
\href{https://link.springer.com/book/10.1007/978-1-4612-2256-9}{Inspire}.

\bibitem{McAvity:1995zd}
D.~M.~McAvity and H.~Osborn,
``Conformal field theories near a boundary in general dimensions,''
Nucl. Phys. B \textbf{455} (1995), 522-576
\href{https://arxiv.org/pdf/cond-mat/9505127.pdf}{[arXiv:cond-mat/9505127 [cond-mat]]}.

\bibitem{He:2020udl}
S.~He and Y.~Sun,
``Correlation functions of CFTs on a torus with a $T\overline{T}$ deformation,''
Phys. Rev. D \textbf{102}, no.2, 026023 (2020)
\href{https://arxiv.org/pdf/2004.07486.pdf}{[arXiv:2004.07486  [hep-th]]}.

\bibitem{Bhattacharya:2012mi}
J.~Bhattacharya, M.~Nozaki, T.~Takayanagi and T.~Ugajin,
``Thermodynamical Property of Entanglement Entropy for Excited States,''
Phys. Rev. Lett. \textbf{110} (2013) no.9, 091602
\href{https://arxiv.org/pdf/1212.1164.pdf}{[arXiv:1212.1164 [hep-th]]}.

\bibitem{He:2020cxp}
S.~He, Y.~Sun and Y.~X.~Zhang,
``$T \overline{T} $-flow effects on torus partition functions,''
JHEP \textbf{09}, 061 (2021)
\href{https://arxiv.org/pdf/2011.02902.pdf}{[arXiv:2011.02902 [hep-th]]}.

\bibitem{Calabrese:2007rg}
P.~Calabrese and J.~Cardy,
``Quantum Quenches in Extended Systems,''
J. Stat. Mech. \textbf{0706}, P06008 (2007)
\href{https://arxiv.org/pdf/0704.1880.pdf}{[arXiv:0704.1880 [cond-mat.stat-mech]]}.

\bibitem{Li:2020pwa}
Y.~Li and Y.~Zhou,
``Cutoff AdS$_{3}$ versus $ T\overline{T} $ CFT$_{2}$ in the large central charge sector: correlators of energy-momentum tensor,''
JHEP \textbf{12}, 168 (2020)
\href{https://arxiv.org/pdf/2005.01693.pdf}{[arXiv:2005.01693  [hep-th]]}.

\bibitem{Kraus:2018xrn}
P.~Kraus, J.~Liu and D.~Marolf,
``Cutoff AdS$_{3}$ versus the $ T\overline{T} $ deformation,''
JHEP \textbf{07}, 027 (2018)
\href{https://arxiv.org/pdf/1801.02714 .pdf}{[arXiv:1801.02714   [hep-th]]}.

\bibitem{Calabrese:2005in}
P.~Calabrese and J.~L.~Cardy,
``Evolution of entanglement entropy in one-dimensional systems,''
J. Stat. Mech. \textbf{0504} (2005), P04010
\href{https://arxiv.org/pdf/cond-mat/0503393.pdf}{[arXiv:cond-mat/0503393 [cond-mat]]}.

\bibitem{Davies:1976hi}
P.~C.~W.~Davies and S.~A.~Fulling,
``Radiation from a moving mirror in two-dimensional space-time conformal anomaly,''
\href{https://www.jstor.org/stable/79130}{Proc. Roy. Soc. Lond. A \textbf{348}, 393-414 (1976)}.

\bibitem{Hawking:1974rv}
S.~W.~Hawking,
``Black hole explosions,''
\href{https://www.quantamagazine.org/black-hole-paradoxes-reveal-a-fundamental-link-between-energy-and-order-20200528}{Nature \textbf{248} (1974), 30-31}.

\bibitem{Hawking:1975vcx}
S.~W.~Hawking,
``Particle Creation by Black Holes,''
\href{https://link.springer.com/article/10.1007/BF02345020#additional-information}{Commun. Math. Phys. \textbf{43} (1975), 199-220}.

\bibitem{Reyes:2021npy}
I.~A.~Reyes,
``Moving Mirrors, Page Curves, and Bulk Entropies in AdS2,''
Phys. Rev. Lett. \textbf{127}, no.5, 051602 (2021)
\href{https://arxiv.org/pdf/2103.01230 .pdf}{[arXiv:2103.01230   [hep-th]]}.

\bibitem{Akal:2020twv}
I.~Akal, Y.~Kusuki, N.~Shiba, T.~Takayanagi and Z.~Wei,
``Entanglement Entropy in a Holographic Moving Mirror and the Page Curve,''
Phys. Rev. Lett. \textbf{126} (2021) no.6, 061604
\href{https://arxiv.org/pdf/2011.12005.pdf}{[arXiv:2011.12005 [hep-th]]}.

\bibitem{Bianchi:2022ulu}
L.~Bianchi, S.~De Angelis and M.~Meineri,
``Radiation, entanglement and islands from a boundary local quench,''
\href{https://arxiv.org/pdf/2203.10103.pdf}{[arXiv:2203.10103 [hep-th]]}.

\bibitem{NIK}
N. I.~Akhiezer, Elements of the Theory of Elliptic
Functions (American Mathematical Society, Providence,
1990).


\end{thebibliography}
\end{document}